\documentclass{article}
\usepackage{graphicx}
\usepackage{hyperref}
\usepackage{amssymb}
\usepackage{amsmath}
\usepackage{amsfonts}
\usepackage{parskip}
\usepackage{titling}
\usepackage{multirow}
\usepackage{dcolumn}
\usepackage[usenames, dvipsnames]{color}
\usepackage{empheq}
\usepackage{xspace}
\usepackage{upgreek}
\usepackage{soul}

\pretitle{\Large\bf} \posttitle{\par\vskip 0.3em} \preauthor{\large} \postauthor{\par\vskip 0.0em}
\predate{\large} \postdate{\par\vskip -1.1em \rule{\linewidth}{1pt}} 

\newcommand{\ket}[1]{\left| {#1} \right\rangle}
\newcommand{\beq}{\begin{equation}}
\newcommand{\eeq}{\end{equation}}
\newcommand{\bea}{\begin{eqnarray}}
\newcommand{\eea}{\end{eqnarray}}

\newcommand{\comment}[1]{}

\begin{document}

\newcommand{\dwave}{D-Wave\xspace}

\newcommand{\maxcut}{MAX-CUT\xspace}

\newcommand{\MIT}{1}
\newcommand{\NII}{2}
\newcommand{\NTT}{3}
\newcommand{\SU}{4}
\newcommand{\CU}{5}
\newcommand{\NASA}{6}
\newcommand{\USRA}{7}
\newcommand{\CIT}{8}
\newcommand{\NTTb}{9}
\newcommand{\JST}{10}

\title{Experimental investigation of performance differences between Coherent Ising Machines and a quantum annealer}

\author{Ryan~Hamerly$^{\MIT,\NII,\dagger}$,      
		Takahiro~Inagaki$^{\NTT,\dagger}$,       
		Peter~L.~McMahon$^{\SU,\NII,\CU,\dagger}$,   
		Davide~Venturelli$^{\NASA,\USRA}$,  \\   
		Alireza~Marandi$^{\CIT,\SU}$,                 
		Tatsuhiro~Onodera$^{\SU}$,               
		Edwin~Ng$^{\SU}$,                        
		Carsten~Langrock$^{\SU}$,                
		Kensuke~Inaba$^{\NTT}$,                  
		Toshimori~Honjo$^{\NTT}$,                
		Koji~Enbutsu$^{\NTTb}$,                  
		Takeshi~Umeki$^{\NTTb}$,                 
		Ryoichi~Kasahara$^{\NTTb}$,              
        Shoko~Utsunomiya$^{\NII}$,               
        Satoshi Kako$^{\NII}$,                   
        Ken-ichi Kawarabayashi$^{\NII}$,         
		Robert~L.~Byer$^{\SU}$,                  
		Martin~M.~Fejer$^{\SU}$,                 
        Hideo~Mabuchi$^{\SU}$,                   
        Dirk~Englund$^{\MIT}$,                   
		Eleanor~Rieffel$^{\NASA}$,               
		Hiroki~Takesue$^{\NTT}$,                 
		Yoshihisa Yamamoto$^{\SU,\JST}$          
		}

\date{\today}
\maketitle
\begin{flushleft}
$^{\MIT}$      \textit{Research Laboratory of Electronics, Massachusetts Institute of Technology, 50 Vassar Street, Cambridge, MA 02139, USA} \\
$^{\NII}$      \textit{National Institute of Informatics, Hitotsubashi 2-1-2, Chiyoda-ku, Tokyo 101-8403, Japan} \\
$^{\NTT}$      \textit{NTT Basic Research Laboratories, NTT Corporation, 3-1 Morinosato Wakamiya, Atsugi, Kanagawa 243-0198, Japan} \\
$^{\SU}$       \textit{E.\ L.\ Ginzton Laboratory, Stanford University, Stanford, CA 94305, USA} \\
$^{\CU}$       \textit{School of Applied and Engineering Physics, Cornell University, Ithaca, NY 14853, USA} \\
$^{\NASA}$     \textit{NASA Ames Research Center Quantum Artificial Intelligence Laboratory (QuAIL), Mail Stop 269-1, Moffett Field, California 94035, USA} \\
$^{\USRA}$     \textit{USRA Research Institute for Advanced Computer Science (RIACS), 615 National Ave, Mountain View, California 94035, USA} \\
$^{\CIT}$      \textit{California Institute of Technology, Pasadena, CA 91125, USA} \\
$^{\NTTb}$     \textit{NTT Device Technology Laboratories, NTT Corporation, 3-1 Morinosato Wakamiya, Atsugi, Kanagawa 243-0198, Japan} \\
$^{\JST}$      \textit{ImPACT Program, Japan Science and Technology Agency, Gobancho 7, Chiyoda-ku, Tokyo 102-0076, Japan} \\
$^\dagger$     These authors contributed equally to this work.
\end{flushleft}

\begin{abstract}
Physical annealing systems provide heuristic approaches to solving 
NP-hard Ising optimization problems.  Here, we study the performance of two types of annealing machines---a commercially available quantum annealer built by \dwave Systems,  and measurement-feedback coherent Ising machines (CIMs) based on optical parametric oscillator networks---on two classes of problems, the Sherrington-Kirkpatrick (SK) model and MAX-CUT. The D-Wave quantum annealer outperforms the CIMs on MAX-CUT on regular graphs of degree 3. On denser problems, however, we observe an exponential penalty for the quantum annealer ($\exp(-\alpha_\textrm{DW} N^2)$) relative to CIMs ($\exp(-\alpha_\textrm{CIM} N)$) for fixed anneal times, on both the SK model and on 50\%-edge-density MAX-CUT, where the coefficients $\alpha_\textrm{CIM}$ and $\alpha_\textrm{DW}$ are problem-class-dependent. On instances with over $50$ vertices, a several-orders-of-magnitude time-to-solution difference exists between CIMs and the \dwave annealer.  An optimal-annealing-time analysis is also consistent with a significant projected performance difference.  The difference in performance between the sparsely connected D-Wave machine and the measurement-feedback facilitated all-to-all connectivity of the CIMs provides strong experimental support for efforts to increase the connectivity of quantum annealers.
\end{abstract}

\section*{Introduction}

Optimization problems are ubiquitous in science, engineering, and business.  Many important problems (especially combinatorial problems such as scheduling, resource allocation, route planning or community detection) belong to the NP-hard complexity class, and even for typical instances require a computation time that scales exponentially with the problem size \cite{Rudich2004}.  Canonical examples such as Karp's 21 NP-complete problems \cite{Karp1972} have attracted much attention from researchers seeking to devise new optimization methods, because by definition any NP-complete problem can be reduced to any other problem in NP with only polynomial overhead.  Many approximation algorithms and heuristics (e.g., relaxations to semidefinite programs \cite{GW1995}, simulated annealing \cite{Kirkpatrick1983}, and breakout local search \cite{Benlic2013}) have been developed to search for good-quality approximate solutions as well as ground states for sufficiently small problem sizes. However, for many NP-hard optimization problems, even moderately sized problem instances can be time-consuming to solve exactly or even approximately.  Hence, there is strong motivation to find alternative approaches that can consistently beat state-of-the-art algorithms. 

Despite decades of Moore's Law scaling, large NP-hard problems remain very costly even on modern microprocessors. Thus, there is a growing interest in special-purpose machines that implement a solver directly by mapping the optimization to the underlying physical dynamics.  Examples include digital CMOS annealers \cite{Yoshimura2015, Tsukamoto2017}, as well as analog devices such as nano-magnet arrays \cite{Sutton2017}, electronic oscillators \cite{Parihar2017, Yin2018} and laser networks \cite{Tait2017}.  Quantum adiabatic computation \cite{Farhi2001} and quantum annealing \cite{Kadowaki1998, Santoro2006, Boixo2014, Santoro2002} are also prominent examples, and may offer the possibility of quantum speedup \cite{Somma2012, Ronnow2014, Albash2018} for certain NP-hard problems.  However, all the non-photonic analog optimization systems realized to date suffer from limited connectivity, so that actual problems must in general first be {\it embedded} \cite{Choi2008, Cai2014} into the solver architecture native graph before they can be solved.  This requirement adds an upfront computational cost \cite{Choi2008, Choi2011, Klymko2014} of finding the embedding (unless previously known) and, of most relevance in this study, in general results in the use of multiple physical pseudo-spins to encode each logical spin variable, which can lead to a degradation of time-to-solution.

In this paper, we perform the first direct comparison between the \dwave 2000Q quantum annealer and the Coherent Ising Machine (CIM) \cite{McMahon2016, Inagaki2016}. As we will see later, a crucial distinction between these systems is their intrinsic connectivity, which has a profound influence on their performance.  Both systems are designed to solve the classical Ising problem, that is, to minimize the classical Hamiltonian:
\beq
	H = \frac{1}{2} \sum_{ij} J_{ij} \sigma_i \sigma_j + \sum_i h_i \sigma_i
\label{eq:1}
\eeq
where $\sigma_i = \pm 1$ are the Ising spins, $J_{ij}$ are the entries of the spin-spin coupling matrix, and $h_i$ the Zeeman (bias) terms.  The Ising problem is NP-hard for non-planar couplings \cite{Barahona1982} and is one of the most widely studied problems in this complexity class.  We focus on two canonical NP-hard Ising problems: unweighted \maxcut \cite{Karp1972} and ground-state computation of the Sherrington-Kirkpatrick spin-glass model \cite{Kirkpatrick1975}.

\begin{figure}[tb]
\begin{center}
\includegraphics[width=1.00\textwidth]{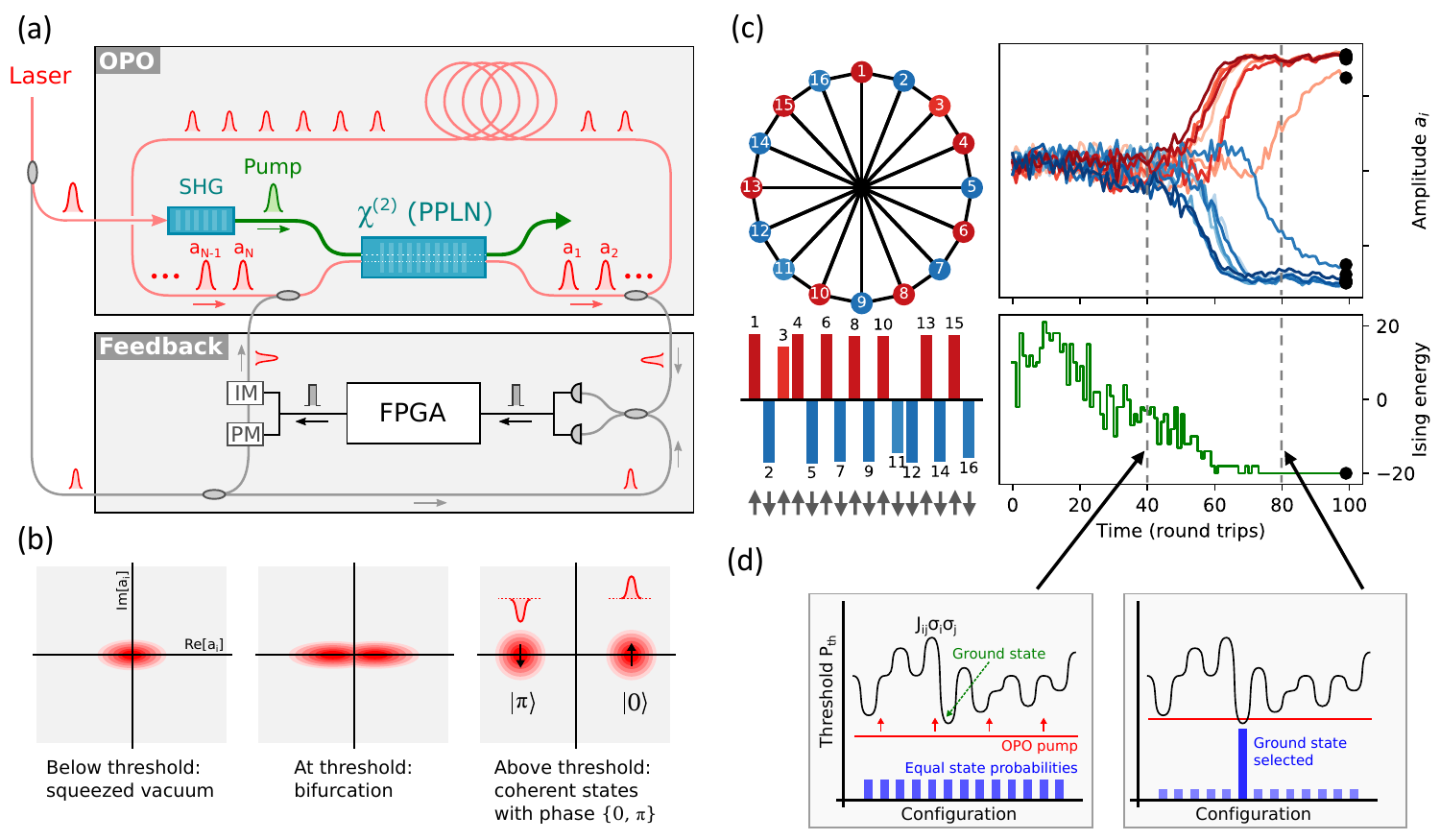}
\caption{(a) Coherent Ising Machine design consisting of time-multiplexed OPO and measurement-feedback apparatus. See Refs. \cite{McMahon2016,Inagaki2016} for details. (b) OPO state during transition from below-threshold squeezed state to (bistable) above-threshold coherent state.  (c) Solution of antiferromagnetic Ising problem on the M\"obius ladder with the CIM, giving measured OPO amplitudes $a_i$ and Ising energy $H$ as a function of time in round trips.  (d) Illustration of search-from-below principle of CIM operation.}
\label{fig:f1}
\end{center}
\end{figure}

In the CIM, the spin network is represented by a network of degenerate optical parametric oscillators (OPOs).  Each OPO is a nonlinear oscillator that converts pump light to its half-harmonic \cite{Boyd2003}; it can oscillate in two identical phase states, which encode the value of the Ising spin \cite{Wang2013, Marandi2014}.  Optical coherence is essential to the CIM, where the data is encoded in the phase of the light.  As Fig.~\ref{fig:f1}(a) shows, time multiplexing offers a straightforward way to generate many identical OPOs in a single cavity \cite{Marandi2014}.  A pulsed laser with repetition time $T$ is used to pump an optical cavity with round-trip time $N\times T$.  Parametric amplification is provided by the $\chi^{(2)}$ crystal; since this is an instantaneous nonlinearity, the circulating pulses in the cavity are identical and non-interacting.  The approach is scalable using high repetition-rate lasers and long fiber cavities: OPO gain has been reported for up to $N = 10^6$ pulses, and stable operation achieved for $N = 50\mbox{,}000$ \cite{Takesue2016}.  Each circulating pulse represents an independent OPO with a single degree of freedom $a_i$.  Classically, $a_i$ is a complex variable, which maps to the annihilation operator $\hat{a}_i$ in quantum mechanics \cite{Walls2007}. A measurement-feedback apparatus is used to apply coupling between the pulses \cite{McMahon2016, Inagaki2016}.  In each round trip, a small fraction of the light (${\sim}10\%$) is extracted from the cavity and homodyned against a reference pulse (the OPO pump is created from second harmonic generation (SHG) of the reference laser, so there is good matching between the reference and the OPO signal light, which is at half the frequency of the pump).  The homodyne result, in essence a measurement of $a_i$, is fed into an electronic circuit (consisting of an ADC, an FPGA, and a DAC) that, for each pulse, computes a feedback signal that is proportional to the matrix-vector product $\sum_j J_{ij} a_j$.  This signal is converted back to light using an optical modulator and a reference pulse, and re-injected into the cavity.  The measurement-feedback CIM has intrinsic all-to-all connectivity through its exploitation of memory in the electronic circuit (although the same effect can be obtained with optical delay-line memories in all-optical CIMs \cite{Wang2013,Marandi2014}). 

The OPO is a dissipative quantum system with a pitchfork bifurcation well adapted for modeling Ising spins: as the pump power is increased (Fig.~\ref{fig:f1}(b)), the OPO state transitions from a below-threshold squeezed vacuum state \cite{Milburn1981, Drummond1981, Wu1986} to an above-threshold coherent state \cite{Kinsler1991}.  Because degenerate parametric amplification is phase-sensitive, only two phase states are stable above threshold; thus the OPO functions as a classical ``spin'' with states $\bigl\{\ket{0}, \ket{\pi}\bigr\}$ that can be mapped to the Ising states $\sigma_i = \{+1, -1\}$.  The optimization process happens in the near-threshold regime where the dynamics are determined by a competition between the network loss and Ising coupling (which seek to minimize the product $\sum_{ij} J_{ij} a_i a_j$), and nonlinear parametric gain (which seeks to enforce the constraints $a_i \in \mathbb{R}, |a_i| = \mbox{const}$).

As an example, consider the Ising problem on the $N = 16$ M\"obius ladder graph with anti-ferromagnetic couplings \cite{Takata2016}.  Fig.~\ref{fig:f1}(c) shows a typical run of the CIM, resulting in a solution that minimizes the Ising energy (data from Ref.~\cite{McMahon2016}).  The most obvious interpretation of the process is spontaneous symmetry breaking of a pitchfork bifurcation: prepared in a squeezed vacuum state and driven by shot noise, the OPO state bifurcates, during which its amplitudes $a_i$ grow either in positive or negative value, and subsequently the system settles into the Ising ground state (or a low-lying excited state) \cite{McMahon2016,Hamerly2016} (this is related to the Gaussian-state model in Ref.~\cite{Clements2017}).  Another view derives from ground-state ``search from below'' (Fig.~\ref{fig:f1}(d)).  Here the Ising energy is visualized as a complicated landscape of potential oscillation thresholds, each with its own spin configuration.  If the OPO pump is far below the minimum threshold, all spin configurations will be excited with near-equal probability, but once the ground-state threshold is exceeded, its probability will grow exponentially at the expense of other configurations \cite{Marandi2014, Yamamoto2017}.  This ground-state selection process corresponds to the $40 \leq t \leq 60$ region in Fig.~\ref{fig:f1}(c).

\begin{figure}[b!]
\begin{center}
\includegraphics[width=1.00\textwidth]{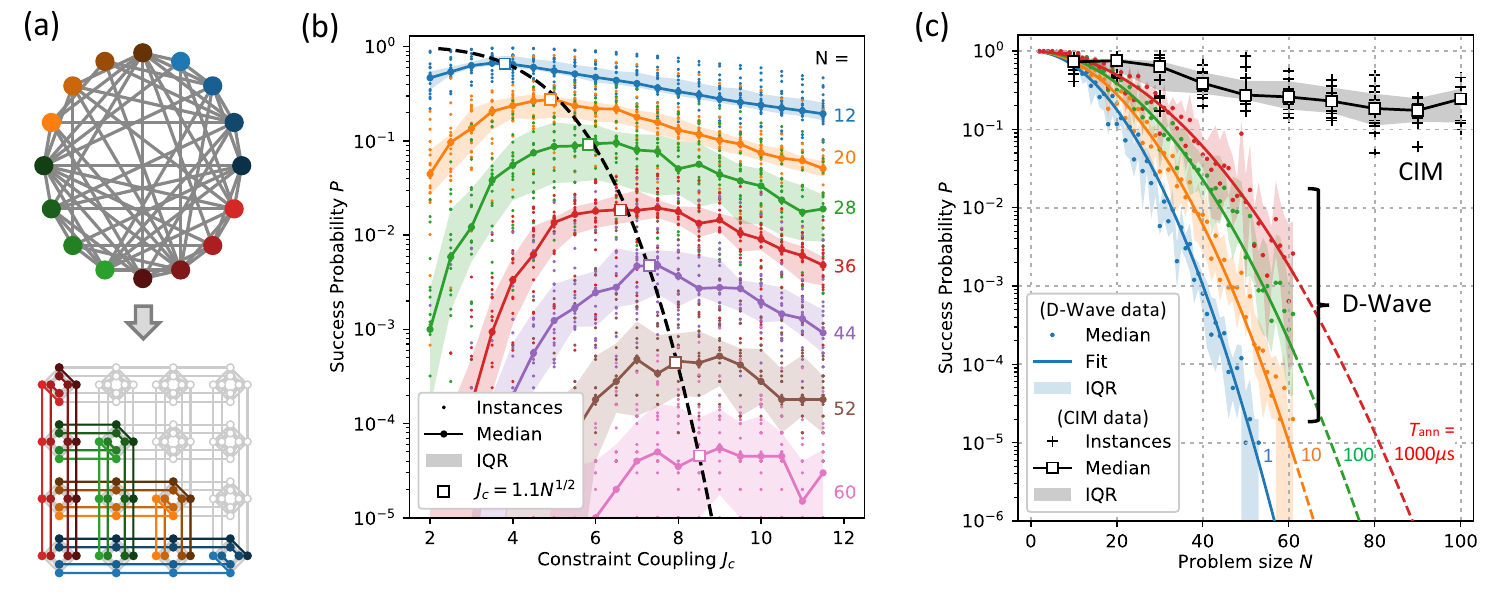}
\caption{(a) Illustration of clique embedding: an arbitrary $N = 16$ graph is embedded into the \dwave chimera, each spin mapped to a ferromagnetically coupled line of physical qubits (each color is a logical qubit).  (b) \dwave ground-state probability for Sherrington-Kirkpatrick (SK) model as a function of problem size $N$ and embedding parameter $J_c$ (annealing time $T_{\rm ann} = 20\upmu$s).  Shading indicates interquartile range (IQR, 25/75 percentile range of instances).  (c) Scaling of ground-state probability for DW2Q (with optimal $J_c$) and Stanford CIM.  \dwave and CIM ran 20 and 10 instances per problem size, respectively.
}
\label{fig:f2}
\end{center}
\end{figure}

The \dwave 2000Q (DW2Q) quantum annealer used in this work is installed at NASA Ames Research Center in Mountain View, California. The DW2Q has 2,048 qubits, but its ``Chimera'' coupling graph (i.e., the graph whose edges define the non-zero $J_{ij}$ terms in Eq.~(\ref{eq:1})) is very sparse.  Since most Ising problems are not defined on subgraphs of the Chimera, \emph{minor embedding} is used to find a Chimera subgraph on which the corresponding Ising model has a ground state that corresponds to the classical ground state of the Ising model defined on the desired problem graph \cite{Choi2008, Cai2014}.  Native clique embeddings \cite{Boothby2016} (Fig.~\ref{fig:f2}(a)) are pre-computed embeddings that can be used for fully-connected problems or problems on dense graphs.  Each {\it logical} qubit is associated to an L-shaped ferromagnetic chain of $\lceil N/\kappa\rceil+1$ {\em{physical}} qubits, where $2\kappa$ is the number of qubits in each unit cell of the Chimera graph ($\kappa$ = 4 in the \dwave 2000Q).  Clique embeddings are desirable because all chain lengths are equal: this architecture simplifies the parameter setting procedure due to symmetry and it is thought to prevent desynchronized freeze-out of chains during the calculation \cite{Venturelli2015}.  However, the embedding introduces considerable overhead relative to the fully-connected model: for $N$ logical qubits, $N (\lceil N/\kappa\rceil + 1) \approx N^2/\kappa$ physical qubits are used.  Due to the triad structure \cite{Choi2011} of the embeddings (Fig.~\ref{fig:f2}(a)), only approximately half of the annealer's physical qubits are utilized, limiting the \dwave 2000Q to problems with $N \leq 64$ (the actual limit is $N \leq 61$ due to unusable qubits on the particular machine at NASA Ames).

\section*{Results}

\subsection*{Sherrington-Kirkpatrick (SK) spin-glass}

As a first benchmarking problem, we consider the Sherrington-Kirkpatrick (SK) spin-glass model on a fully-connected (i.e., maximally dense) graph, where the couplings $J_{ij} = \pm 1$ are randomly chosen with equal probability \cite{Kirkpatrick1975}.  Ground-state computation of the SK model is directly related to the graph partitioning problem, which is also NP-hard \cite{Fu1986}.  For each problem size $2 \leq N \leq 61$, 20 randomly-chosen instances were solved on the DW2Q.  We consider as a performance metric the success probability $P$, defined as the fraction of runs on the same instance that return the ground state energy (we will discuss the time to solution, which requires a more thorough analysis involving optimal annealing time, in a subsequent section).

To properly benchmark the DW2Q's performance, it is important to optimize the embedding parameter $J_c$, which sets the ratio of constraint couplings to logical couplings.  The optimal $J_c$ depends on the problem type and size and is found empirically.  The DW2Q success probability is plotted as a function of $J_c$ in Fig.~\ref{fig:f2}(b), from which we find that the optimal $J_c$ scales roughly as $N^{1/2}$ (see methods for details).  This scaling is consistent with results published on the same class of problems with the earlier D-Wave Two quantum annealer, and it is believed to be connected to the spin-glass nature of the SK Ising problem \cite{Venturelli2015}.  Using the optimal value of $J_c$, Fig.~\ref{fig:f2}(c) shows that the performance on the \dwave depends strongly on the single-run annealing time, with the values $T_{\rm ann} = (1, 10, 100, 1000)\upmu \textrm{s}$ plotted here.  The \dwave annealing time is restricted to the range $[1, 2000]\upmu \textrm{s}$.  We observe that longer annealing times give higher success probabilities, in accordance with the expectations from the adiabatic quantum optimization approach that inspired the design of the \dwave machine.  The data fit well to a square-exponential $P = \exp(-(N/N_0^\textrm{DW})^2)$, where the parameter $N_0^\textrm{DW}$ increases slowly, roughly logarithmically, with $T_{\rm ann}$.  For problem sizes $N<30$, the results in Fig.~\ref{fig:f2}(c) agree with an extrapolation of the benchmark data for $T_{\rm ann}=20\upmu{\rm s}$ reported in Ref.~\cite{Venturelli2015}, which used an earlier processor (the 512-qubit D-Wave, despite the engineering improvements that have been made in the last two generation chips (2X and 2000Q)).

The same SK instances for $N = 10, 20, \ldots, 60$ were solved on the CIMs hosted at Stanford University in Stanford, California and NTT Basic Research Laboratories in Atsugi, Japan \cite{McMahon2016, Inagaki2016}.  Additional problems with $N \geq 60$ were also solved on the CIM, but were too large to be programmed on the DW2Q.  The two Ising machines have similar performance (see Supp.\ Sec.~\ref{sec:supp2} for more details).  Fig.~\ref{fig:f2}(c) shows a plot of the success probability as a function of problem size: the exponential scaling for the CIM is shallower than the one given by the DW2Q performance. We note that the success probability $P$ for the CIM scales approximately as $\exp(-N/N_0^\textrm{CIM})$, where $N_0^\textrm{CIM}$ is a constant. The fact that for the DW2Q, success probability $P$ scales with an $N^2$ dependence in the exponential rather than $N$ (as is the case for the CIM) leads to a dramatic difference in success probability between the quantum annealer and the CIM for problem sizes $N \ge 60$.

\subsection*{\maxcut}

We next study the DW2Q performance on \maxcut for both dense and sparse unweighted graphs.  Unweighted \maxcut is the problem of finding a partition (called a cut) of the vertices $V$ of a graph $G = (V, E)$ where the partition is defined by two disjoint sets $V_1$ and $V_2$ with  $V_1 \cup V_2 = V$, and for which the number of edges between the two sets $\left|\left\{(v_1 \in V_1, v_2 \in V_2) \in E\right\}\right|$ is maximized.  Unweighted \maxcut is NP-hard for general graphs \cite{Karp1972}, and can be expressed as an Ising problem by setting the anti-ferromagnetic couplings $J_{ij} = +1$ along graph edges: $H = \sum_{(ij)\in E} \sigma_i\sigma_j$.  Thus, the problem in Fig.~\ref{fig:f1}(c) is the same as \maxcut on the M\"obius ladder graph.  Previous CIM studies have solved \maxcut on problems up to size $N = 2\mbox{,}000$ in experiment \cite{Marandi2014, Takata2016, McMahon2016, Inagaki2016} and $N = 20\mbox{,}000$ in simulation \cite{Wang2013, Haribara2016}.

\begin{figure}[tbp]
\begin{center}
\includegraphics[width=1.00\textwidth]{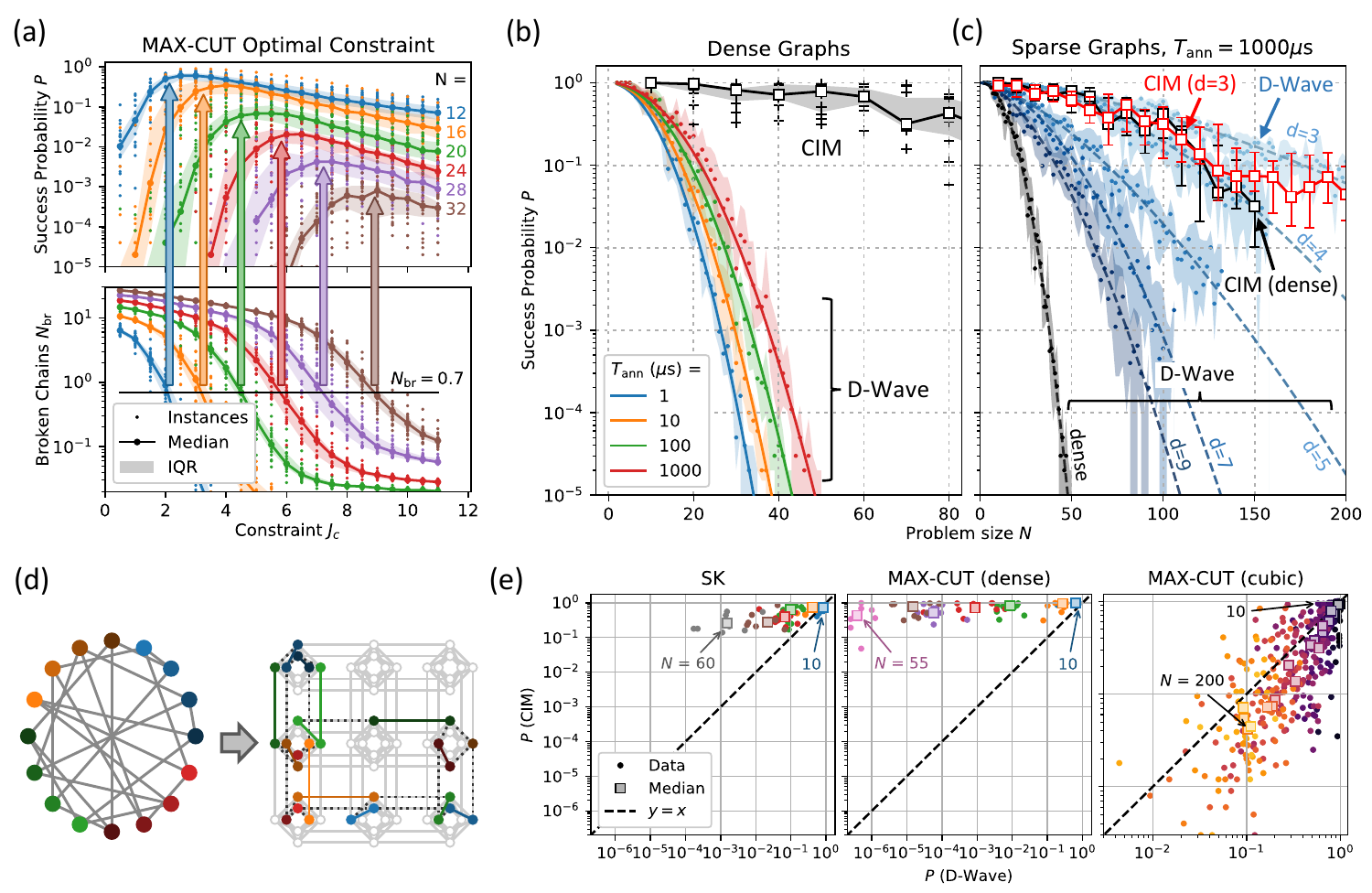}
\caption{(a) \dwave performance on dense \maxcut problems with edge density 0.5, showing that optimal performance occurs when the $J_c$ coupling is strong enough to make it unlikely that logical qubits (chains) become ``broken'' (see also Supp.\ Figs.~\ref{fig:fs1}-\ref{fig:fs2}).  (b) \dwave and NTT CIM success probability for dense \maxcut as a function of problem size (for $T_{\rm soln}$ see Supp.\ Fig.~\ref{fig:fs6}). (c) \dwave (annealing time $T_{\rm ann} = 1000\upmu{\rm s}$) and NTT CIM success probability for sparse graphs of degree $d = 3, 4, 5, 6, 9$ as well as dense graphs.  (d) Example of a cubic-graph embedding found with the heuristic.  (e) Success probability scatterplots comparing D-Wave ($T_{\rm ann} = 1000\upmu{\rm s}$) and CIM.
}
\label{fig:f3}
\end{center}
\end{figure}

Random unweighted \maxcut graphs at the phase transition \cite{coppersmith2004random}, with edge density 0.5 (i.e.\ Erd\H{o}s-R\'{e}nyi graphs $G(N,\tfrac{1}{2})$) were tested on DW2Q for problems up to $N = 61$, and on the CIM for $N \leq 150$.    For these graphs, clique embeddings were used, but in practice the performance did not differ from the embedding heuristic provided by the \dwave API \cite{Cai2014}.  In Fig.~\ref{fig:f3}(a) we show that the optimal value of the embedding coupling parameter $J_c$ appears to be correlated with the appearance of defects in the perfect polarization state expected in logical qubits at the end of the anneal.  With $J_c$ optimized, the success probability follows the same square-exponential ($e^{-O(N^2)}$) trend with $N$ as in the SK model, but the drop-off is even steeper.  The CIM success probabilities are also lower than for the SK model, but are now orders of magnitude higher than the DW2Q for $N\ge40$.

To test the effect of sparseness, Fig.~\ref{fig:f3}(c) plots the performance on unweighted regular graphs of degree $d = 3, 4, 5, 7, 9$, where the degree of a graph is the number of edges per vertex.  Despite their sparseness, \maxcut on these restricted graph classes is also NP-hard \cite{Alimonti1997}.  The CIM shows no performance difference between $d = 3$ (cubic) and dense graphs.  For DW2Q, the sparse graphs are embedded using the graph minor heuristic, which allows problems of up to size $N = 200$ to be embedded in the DW2Q \cite{Cai2014}.  In addition, the found embeddings require significantly fewer qubits (for the sparse graphs) than the clique embeddings (compare Figs.~\ref{fig:f3}(d) and \ref{fig:f2}(a), see also Supp.~Fig.~\ref{fig:fs3}(b)).  For cubic graphs, the DW2Q achieves slightly better performance than the CIM, while the CIM's advantage is noticeable for $d \geq 5$.

The CIM achieves similar success probabilities for cubic and dense graphs, suggesting that dense problems are not intrinsically harder than sparse ones for this class of annealer.  \dwave's strong dependence on edge density is most likely a consequence of embedding compactness: it is known that more compact embeddings (fewer physical qubits per chain) tend to give better annealing performance, after all optimization and parameter setting is considered \cite{Cai2014}.  Since qubits on the \dwave chimera graph have at most 6 connections, the minimum chain length is $\ell = \lceil(d-2)/4\rceil$, so embeddings grow less compact with increasing graph degree (see Supp.\ Sec.~\ref{sec:supp3}).  Since degree-1 and degree-2 vertices can be pruned from a graph in polynomial time (a variant of cut-set conditioning \cite{Dechter1986}), $d = 3$ is the minimum degree required for NP-hardness.  Of NP-hard \maxcut instances, Fig.~\ref{fig:f3}(c) suggests that there is only a very narrow region ($d = 3, 4$) where \dwave matches or outperforms the CIM in success probability; for the remainder of the graphs the CIM dominates.

\subsection*{Time to Solution and Optimal Annealing Time}

Although success probability is a helpful metric to understand scaling for fixed $T_{\rm ann}$, the key computational figure of merit is the \emph{time to solution} $T_{\rm soln} = T_{\rm ann} \lceil \log(0.01)/\log(1 - P)\rceil$.  This figure multiplies the expected number of independent runs to solve a problem with 99\% probability with the time of a single run, $T_{\rm ann}$.  When evaluating time to solution of a physical annealer, it is important to optimize as much as possible the run parameters, in particular considering the machine's $T_{\rm soln}$ at the {\it optimal} annealing time.  This avoids a common pitfall of fixed-anneal-time analysis, where if the chosen anneal time is too large, near-flat $T_{\rm soln}$ scaling for small problem sizes gives the illusion of speedup \cite{Ronnow2014, McGeoch2018}.

\begin{figure}[tb]
\begin{center}
\includegraphics[width=1.00\textwidth]{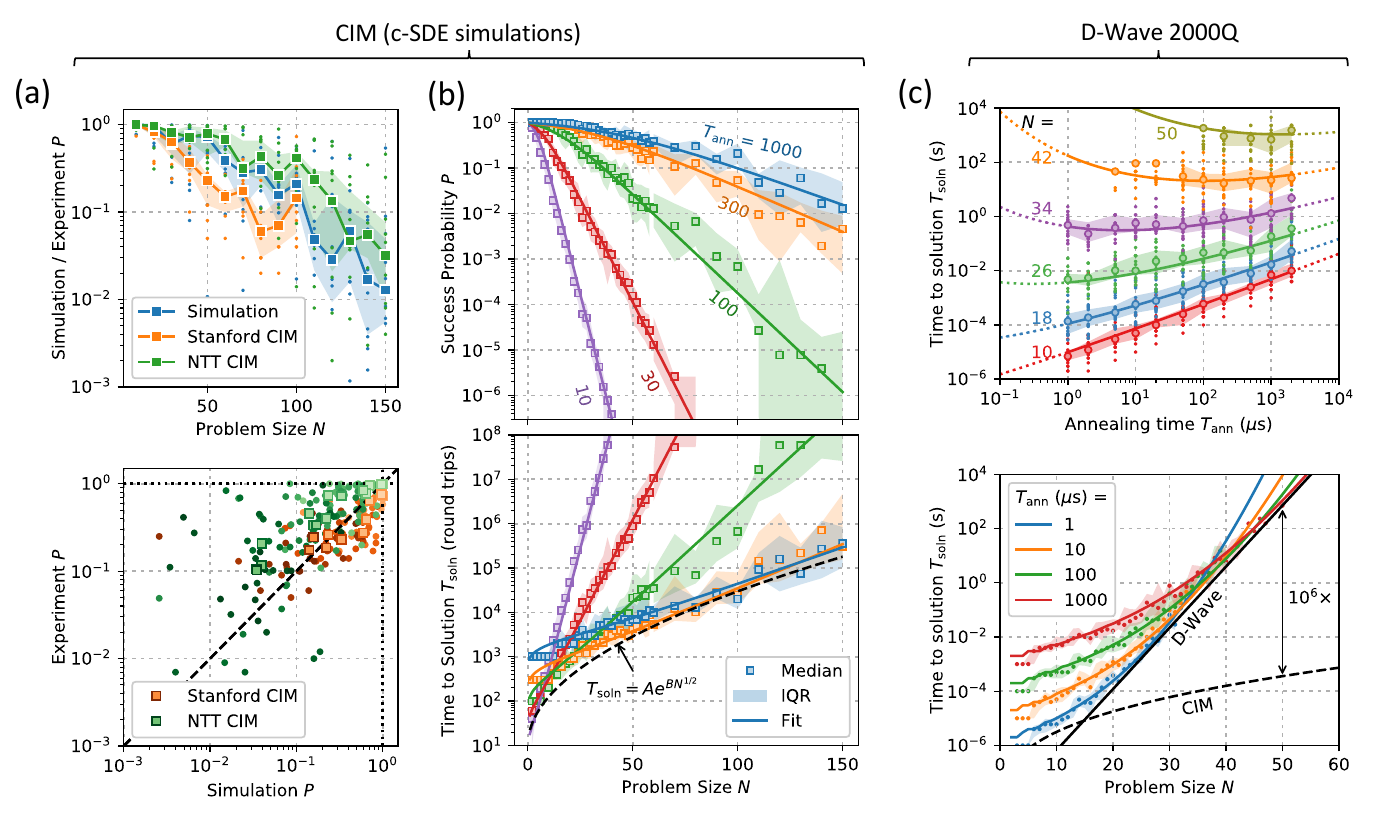}
\caption{Optimal annealing-time analysis for \dwave and CIM, dense \maxcut instances.  (a) CIM experimental performance vs.\ c-number SDE simulations.  (b) CIM success probability and time to solution (given in terms of the number of round trips) as a function of problem size $N$.  The effective round-trip time for the NTT CIM ($(2.5N){\rm ns}$, see Supp.\ Sec.~S2) is used to convert this figure to seconds.  (c) \dwave time to solution as a function of annealing time $T_{\rm ann}$ and problem size $N$.  Dashed line shows optimal CIM $T_{\rm soln}$ from (b) for comparison.}
\label{fig:f4}
\end{center}
\end{figure}

In the CIM, the anneal time is set by the pump turn-on schedule and is an integer number of round trips.  The experiments in this paper were conducted with $T_{\rm ann} = 1000$ round trips, but shorter or longer times are also possible.  To assess the effect of the anneal time on CIM performance, we simulate the CIM with c-number stochastic differential equations (c-SDEs) using the truncated Wigner representation \cite{Kinsler1991}.  The algorithm, which is based on Ref.~\cite{Hamerly2016}, is described in Supp.~Sec.~\ref{sec:supp3}.  Fig.~\ref{fig:f4}(a) compares the performance of the experimental CIMs to the c-SDE model for dense \maxcut instances, indicating that the model reasonably reproduces the behavior of the experimental CIMs (similar agreement is found for SK and cubic \maxcut instances, see Supp.\ Fig.~\ref{fig:fs9}).

Fig.~\ref{fig:f4}(b) plots the (c-SDE) CIM success probability and time to solution (in round trips) as a function of anneal time.  Consistent with the experimental results, we see an exponential scaling with $N$ in the large-$N$ limit (the curves are fit to $P(N) = (a + (1-a) e^{b N})^{-1}$, which becomes exponential for large $N$).  The time-to-solution plot is a series of (nearly) linear intersecting curves, where curves with shorter anneal time have a lower intercept but a larger slope.  Optimizing $T_{\rm soln}$ reveals a tradeoff between the annealing time of a single run and success probability: short anneals are preferred for small problems where the success probability is always close to unity and insensitive to the annealing time, and long anneals are preferred for large problems where the success probability dominates.  Thus the optimal anneal time depends on problem size and increases with $N$.  Fig.~\ref{fig:f4}(c) shows the analogous \dwave data for the same problem class (dense \maxcut).  Here the fixed-$T_{\rm ann}$ curves scale quadratically with $N$ rather than linearly.  The same scaling is observed in SK problems, see Supp.\ Figs.~\ref{fig:fs10}-\ref{fig:fs11}.

It has been observed empirically on \dwave quantum annealers that for Chimera-graph spin glasses, optimal time-to-solution scales as $T_{\rm soln} \propto \exp(O(N^{1/2}))$ \cite{Ronnow2014,Albash2018,Mandra2018}, while $T_{\rm soln}$ at fixed anneal times increases more steeply \cite{Ronnow2014}.  The lower envelope of the curves in Fig.~\ref{fig:f4}(b) can be reasonably fit to this form, suggesting a similar scaling for the CIM, even though the CIM is based on an entirely different computational principle.  Moreover, when solving embedded problems in quantum annealers, the optimal embedding parameters are believed to be associated with the emergence of a spin-glass phase of the embedded problem \cite{Venturelli2015}.  Since for dense graphs the embedded problem has $N_{\rm ph} \propto N^2$ qubits, this would suggest a scaling $T_{\rm soln} \propto \exp\bigl(O((N^2)^{1/2})\bigr) = \exp(O(N))$, as shown in Fig.~\ref{fig:f4}(c).  Both the CIM scaling $T_{\rm soln} \propto \exp(O(N^{1/2}))$ (dashed curves in Fig.~\ref{fig:f4}(b-c)) the DW2Q scaling $T_{\rm soln} \propto \exp(O(N))$ (solid curve in Fig.~\ref{fig:f4}(c)) are consistent with the hypothesis that a physical annealer's time to solution (at optimal annealing time) should scale exponentially with $N_{\rm ph}^{1/2}$, where $N_{\rm ph}$ is the number of {\it physical} qubits (or bits) required to encode the Ising problem.

We note that our claims are only suggestive, but not conclusive, of $\exp(O(N_{\rm ph}^{1/2}))$ scaling at the optimal annealing time.  Only for a limited range of problem sizes ($25 \leq N \leq 50$) is the optimal annealing time accessible with the DW2Q, and the data are noisy enough that other curves would also fit the lower envelope.  Thus we caution against na\"{i}vely extrapolating these curves to large problem sizes.  However, a clear scaling advantage for the CIM does exist at {\it measured} problem sizes, a conclusion also observed (Supp.\ Figs.~\ref{fig:fs10}-\ref{fig:fs11}) for SK and cubic (i.e., $d=3$) \maxcut instances (although the DW2Q nonetheless outperforms the CIM in absolute terms for all measured cubic \maxcut problems).

While the optimal-annealing-time analysis is important theoretically, in realistic machines $T_{\rm ann}$ is limited by parameter misspecification, finite temperature, and various noise sources, and cannot be increased arbitrarily.  Therefore it is relevant to consider the achievable performance for practically realizable choices of $T_{\rm ann}$.  Table~\ref{tab:t1} shows the experimental times-to-solution for the NTT CIM (fixed $T_{\rm ann} = 1000$ round trips) and the DW2Q (for range $T_{\rm ann} \in [1, 1000]\upmu$s).  This allows a comparison of the optimal time-to-solution in experimentally accessible parameters, which may be more a useful benchmark for near-term annealing machines.  The DW2Q outperforms the CIM by a factor of 10--100$\times$ at cubic MAX-CUT problems, although this factor shrinks with increasing problem size.  For the SK and dense MAX-CUT problems, on the other hand, the CIM outperforms D-Wave by several orders of magnitude when $N \geq 40$.  For MAX-CUT, by $N=55$, the factor is $10^7$, and extrapolated to $N = 100$, it exceeds $10^{20}$.

\begin{table}[h]
\begin{center}
\begin{tabular}{cccc|cccc|cccc}
\hline\hline
\multicolumn{4}{c|}{\bf SK} & \multicolumn{4}{c|}{\bf \maxcut (dense)} & \multicolumn{4}{c}{\bf \maxcut ($d = 3$)} \\ 
$N$ & DW2Q & CIM & Factor & $N$ & DW2Q & CIM & Factor & $N$ & DW2Q & CIM & Factor \\ \hline
10            & 6.0 $\upmu$s    & 25 $\upmu$s  & 0.2 &
10            & 6.0 $\upmu$s    & 25 $\upmu$s  & 0.2 & 
10            & 1.0 $\upmu$s    & 50 $\upmu$s  & 0.02 \\
20            & 35 $\upmu$s     & 100 $\upmu$s & 0.3   & 
20            & 0.4 ms        & 100 $\upmu$s & 4   & 
20            & 3.0 $\upmu$s    & 100 $\upmu$s & 0.03 \\
40            & 6.1 ms        & 0.4 ms     & 15   & 
40            & 6.1 s         & 0.4 ms     & $10^4$   & 
50            & 12 $\upmu$s     & 0.4 ms     & 0.03 \\
60            & 1.4 s         & 0.6 ms     & 2000   & 
55            & $10^4$ s         & 1.2 ms     & $10^7$   & 
100           & 100 $\upmu$s    & 3.3 ms     & 0.03 \\
80$^\ast$  & ($400$ s)     & 1.8 ms     & ($10^5$)   & 
80$^\ast$  & ($10^{11}$ s) & 1.8 ms     & ($10^{13}$)   & 
150           & 2.8 ms        & 22 ms      & 0.1 \\
100$^\ast$ & ($10^5$ s)    & 3.0 ms     & ($10^7$)   & 
100$^\ast$ & ($10^{19}$ s) & 2.3 ms     & ($10^{21}$)   & 
200           & 11 ms         & 51 ms      & 0.2 \\
\hline\hline
\multicolumn{12}{p{16cm}}{{\small $^\ast$\dwave solution times extrapolated using $P = e^{(N/N_0)^2}$ fits in Figs.~\ref{fig:f2}(c), \ref{fig:f3}(b).  Note that dense problems with $N > 61$ are not embeddable in the DW2Q.}}
\end{tabular}
\caption{Time to solution $T_{\rm soln}$ for SK, dense \maxcut, and $d = 3$ \maxcut problems on \dwave and NTT CIM (see Supp.\ Sec.~\ref{sec:supp2}).  The annealing time for \dwave runs was chosen (in the range [1, 1000]$\upmu$s) to optimize $T_{\rm soln}$ (see Supp.\ Sec.~\ref{sec:supp4}).  All CIM data are for fixed anneal times (1000 round trips).  ``Factor'' refers to the ratio of solution times $T^{\rm (DW)}_{\rm soln} / T^{\rm (CIM)}_{\rm soln}$.}
\label{tab:t1}
\end{center}
\end{table}

\subsection*{Graph Density and Performance}

\begin{figure}[b!]
\begin{center}
\includegraphics[width=0.95\textwidth]{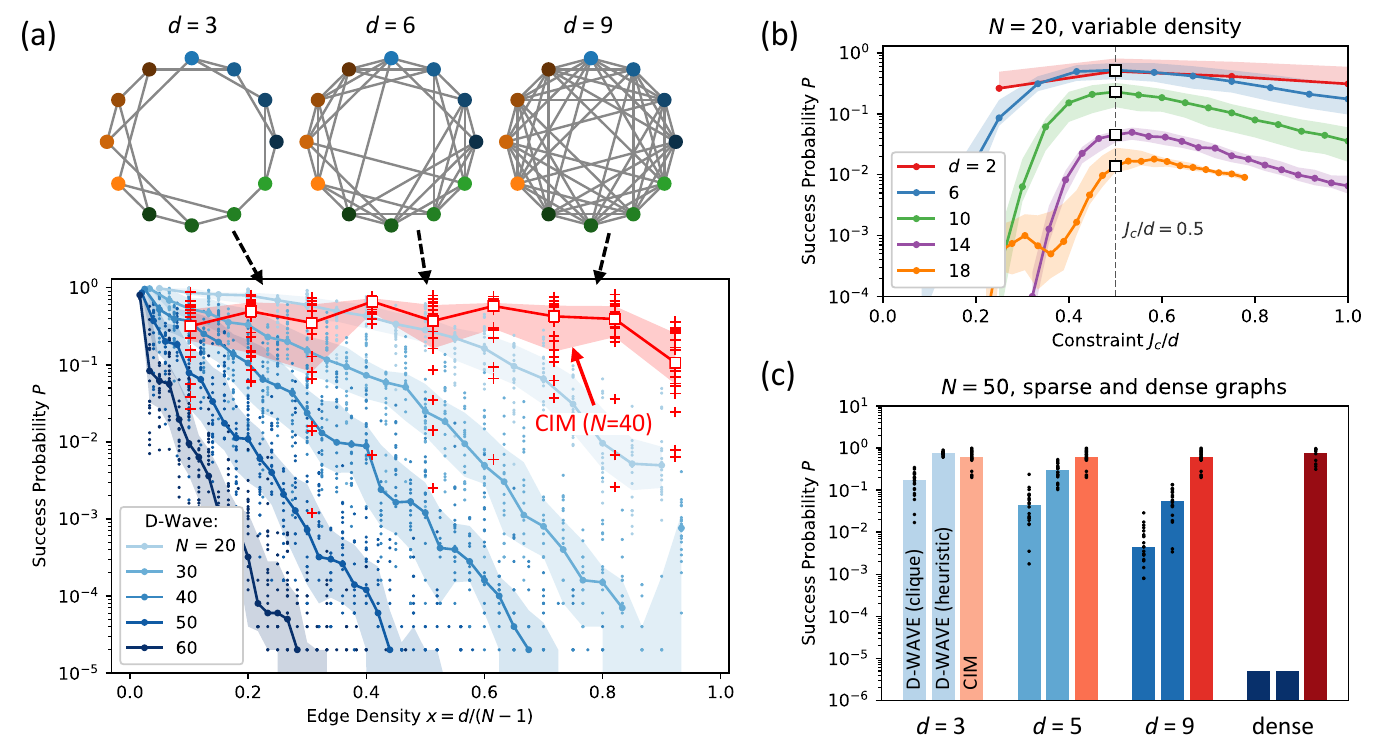}
\caption{(a) Success probability as a function of edge density.  Native clique embeddings used for \dwave.  Optimal embedding parameter (see subgraph (b)) is used, with $T_{\rm ann} = 1000~\upmu$s.  (b) \dwave success probability as a function of graph degree, showing that the optimal $J_c$ scales as $J_c \propto d$ for fixed $N$.  (For fixed edge density, the $N$ dependence was determined previously to be $J_c \propto N^{3/2}$, see Fig.~\ref{fig:f3}(a)).  (c) Comparison of \dwave and NTT CIM success probabilities for $N = 50$, using both clique embeddings and heuristically determined embeddings ($T_{\rm ann} = 1000~\upmu$s, dense \dwave bars are extrapolation from $e^{-(N/N_0)^2}$ fit in Fig.~\ref{fig:f3}(b))
}
\label{fig:f5}
\end{center}
\end{figure}

Earlier in Fig.~\ref{fig:f3}, we compared MAX-CUT on dense graphs and sparse regular graphs to show that the DW2Q's performance depends strongly on graph sparseness.  Fixing the problem size and varying the edge density, we see the same effect and can fill in the gap between sparse graphs and dense graphs.  We constructed random unweighted graphs of degree $d = 1, 2, \ldots, (N-2)$ for each graph size $N = 20, 30, 40, 50, 60$.  The success probabilities for DW2Q and the CIM are shown in Fig.~\ref{fig:f5}(a) (for clarity only $N = 40$ CIM data are shown).  In this case, we used clique embeddings for all problems, so for a given $N$ all the embeddings are the same.  Even with the embeddings fixed, the DW2Q finds sparse problems easier to solve than dense ones.  The reason is that, consistent with Ref.~\cite{Venturelli2015}, the optimal constraint coupling is weaker for sparse problems than for dense problems (Fig.~\ref{fig:f5}(b)).  In general, we find that $J_c \propto d$ for fixed $N$.  Having a large constraint coupling could be problematic because the physical quantum annealer scales the largest coupling coefficient to the maximum coupling strength on the chip; the constraints max out this coupling and cause the logical couplings to be downscaled proportionally as $J_c^{-1}$.  Thus dense graphs have weaker logical couplings in the embedded problem, hindering the annealer's ability to find the ground state due to parameter misspecification or ``intrinsic control errors'' (ICE) \cite{Venturelli2015, DWaveNotes}.

The CIM has only weak dependence on the edge density $x = d/(N-1)$.  Earlier work on $N = 100$ graphs~\cite{McMahon2016}, as well as the CIM data plotted in Fig.~\ref{fig:f3}(c), are consistent with this result.  This suggests that the CIM has promise as a general-purpose Ising solver, achieving good performance on a large class of problems, irrespective of connectivity.

Comparing Figs.~\ref{fig:f5}(a) and \ref{fig:f3}(c) we can glean some insight regarding the effect of embedding overhead on the \dwave quantum annealer's performance.  The heuristic embeddings in Fig.~\ref{fig:f3}(c) are designed to minimize the overhead factor (ratio of physical qubits to logical qubits).  This ratio is much larger for the native-clique embeddings, growing linearly, i.e., as $O(N)$ (see Supp.\ Sec.~\ref{sec:supp1}).  Fig.~\ref{fig:f5}(c) compares these two \dwave settings against the CIM at $N = 50$; while the CIM out-performs on all graphs with $d \geq 5$, the difference between the success probabilities using clique and heuristic embeddings suggests that performance is heavily dependent on embedding overhead and the difference grows with edge density (and graph size).  This illustrates an additional tradeoff in quantum annealing: poor-performing but easy-to-find embeddings vs.\ well-performing embeddings that require substantial pre-computation.  This tradeoff is expected to favor the well-performing embeddings when the number of qubits (or connections) becomes large.

\section*{Discussion}

In conclusion, we have benchmarked the \dwave 2000Q system hosted at NASA Ames and measurement-feedback CIMs hosted at Stanford University and NTT Basic Research Laboratories, focusing on \maxcut problems on random graphs and Sherrington-Kirkpatrick spin-glass models, and found that the merits of each machine are highly problem-dependent.  Connectivity appears to be a key factor in performance differences between these machines.  Problems with sparse connectivity, such as 1D chains (\textit{cf.} Refs.~\cite{Gardas2017} and \cite{Inagaki2016dw}) and \maxcut on cubic graphs (Fig.~\ref{fig:f3}), can be embedded into the DW2Q with little or no overhead, resulting in similar performance between the quantum annealer and the CIMs.  However, the embedding overhead for dense problems like SK is very steep, requiring $O(N^2)$ physical qubits to represent a size-$N$ graph, and resulting in large embedded problems that decrease the performance of the quantum annealer.  The ability to avoid an embedding overhead likely contributes to the CIM's performance advantage on SK models that grows exponentially with the square of the problem size. For problems of intermediate sparseness, such as \maxcut on regular graphs of small degree $d \geq 5$, the CIM is still faster by a large factor. 

Ultimately it is overall quantities such as wall-clock time or energy usage that are of practical interest. Read-in and read-out times, classical pre- and post-processing, and energy usage must be included in a comprehensive evaluation. Both CIMs and superconducting qubit quantum annealers are in early stages of development, with these quantities currently in flux. Moreover, to beat state-of-the-art classical techniques (Supp.\ Sec.~\ref{sec:supp5}) on the problems studied in this paper, advances will be required. D-Wave has recently implemented features allowing unconventional control of the annealing process that can significantly improve results, and as mentioned above, efforts to improve the connectivity are underway.  A key question will be the extent to which these technologies can harness quantum effects for computational purposes. Signatures of entanglement have been seen in D-Wave quantum annealers, though it remains open the extent to which the computation makes use of entanglement-related effects.  CIMs, already interesting as semiclassical computational devices, can in principle also have entanglement \cite{Maruo2016}, by either building scalable all-optical couplings \cite{Marandi2014,shen2017deep} (albeit with low losses being required), or by creating entanglement in the measurement-feedback architecture, for example by performing entanglement swapping.

While the path forward for designing improved CIMs and quantum annealers involves many different aspects, this paper has primarily observed results that can be interpreted as being related to connectivity differences between the machines that were benchmarked. It has been conjectured often that increased internal connectivity in quantum annealers can improve performance \cite{Katzgraber2014, Rieffel2015}, and there are large projects underway to realize higher-connectivity quantum annealers (including efforts by \dwave, as well as the IARPA QEO program \cite{Weber2018}, and Google \cite{Chen2017}). Our results provide strong experimental justification for this line of development.

\section*{Methods}

\subsection*{Sample Problems}

For fully-connected SK and \maxcut on dense graphs, 20 random instances were created of each size $N = 2, 3, \ldots, 61$ for the \dwave.  Of these, the $N = 2, 10, 20, \ldots, 60$ instances were also used for the CIM.  An additional set of random instances were created for $N = 70, 80, \ldots, 150$ for the CIM, using the same algorithm.

For the sparse-graph analysis, we computed regular graphs of size $N = 2, 4, \ldots, 300$ and degree $d = 3, 4, \ldots 20$, with 20 instances for each pair $(N, d)$.  The algorithm randomly assigns edges to eligible vertices until all reach the required degree (and backtracks if it gets stuck).  The same algorithm was also used for the variable-density graphs: $d = 1, 2, \ldots, (N-2)$ for $N = 20, 30, 40, 50, 60$, creating 20 instances per pair $(N, d)$.

Exact SK ground states were found with the Spin Glass Server \cite{SGS}, which uses BiqMac \cite{RRW10}, an exact branch-and-bound algorithm.  For SK instances of size $N \leq 100$, the algorithm obtained proven ground states.  For $N > 100$ the solver timed out before exhausting all branches (runtime $T = 3000\,{\rm s}$), so the result is not a guaranteed ground state; however, we believe it reaches the ground state with high probability for $N \leq 150$ because multiple runs of the algorithm give the same state energy, and none of the CIM runs found an Ising energy lower than the Spin Glass Server result.  \maxcut ground states for $N \leq 30$ were found by brute-force search on a GPU; for $20 \leq N \leq 150$ a Breakout Local Search (BLS) algorithm was used \cite{Benlic2013}.  Although BLS is a heuristic solver, for $N \leq 150$ it finds the ground state with nearly 100\% probability, giving us high confidence that the BLS solutions are ground states.  While the brute-force solver, \dwave, and the CIM found states of equal energy to the BLS solution (if run long enough), they never found states of lower energy.

\subsection*{\dwave annealers}

Initial \dwave experiments were performed on the \dwave 2X at NASA Ames Research Center and the \dwave 2X online system at D-Wave Systems Inc.  Later runs were made on the \dwave 2000Q at NASA Ames, once that machine came online.  The 2X and 2000Q systems use a C12 ($12\;{\rm cells} \times 12\;{\rm cells} \times 4\;{\rm qubits}$) and C16 ($16 \times 16 \times 4$) Chimera, respectively.  For all-to-all graphs, \dwave 2X supports $N \leq 48$ and 2000Q supports $N \leq 64$ (the number is slightly smaller because of broken qubits).  All $N \leq 48$ runs were consistent across the three machines as well as with extrapolation of data in Ref.~\cite{Venturelli2015} from runs performed on a different set of instances on the earlier generation machine \dwave Two.  All data reported in this paper came from the \dwave 2000Q.

Embeddings were pre-computed for all problems (heuristic embeddings {} for sparse \maxcut; native clique embeddings for SK, dense \maxcut, and variable-density \maxcut) so that runs in different conditions (e.g.\ annealing times, constraint couplings) would use the same embeddings.  For each problem type, the optimal annealing parameter $J_c$ is found as a function of problem size $N$ by sweeping $J_c$ (Supp.\ Sec.~\ref{sec:supp1}).  The optimal $J_c$ was found to be independent of the annealing time.  The standard annealing schedule was used in all experiments, but the annealing time was tuned.  Most instances were run $10^4$--$10^5$ times total, depending on the observed success rate (the especially hard $N \geq 50$ \maxcut instances were run up to $4\times 10^6$ times).  5--10 different embeddings were used per instance and the success probability was averaged.  Spin-reversal transformations were used to avoid spurious effects.  After an anneal, each logical qubit value was determined by taking the majority vote of all qubits in the chain.

In all figures, the shaded regions give the $[25, 75]$-percentile range (inter-quartile range, or IQR) for the data.  Figs.~\ref{fig:f2}(b), \ref{fig:f3}(a), \ref{fig:f5}(a), show individual instances as dots and the solid line gives the median.  Figs.~\ref{fig:f2}(c), \ref{fig:f3}(b-c), \ref{fig:f5}(b) are too crowded to show \dwave instances; the dots give medians and the smooth lines give analytic fits.  For CIM data, medians and IQR are shown in Figs.~\ref{fig:f2}(c), \ref{fig:f3}(b), while Fig.~\ref{fig:f3}(c) only shows medians and IQR, due to crowding.

\subsection*{CIM}

CIM experiments were performed on the 100-OPO CIM at Ginzton Laboratory of Stanford University and the 2048-OPO CIM at NTT Basic Research Laboratories. The Stanford and NTT devices are described in Refs.~\cite{McMahon2016} and \cite{Inagaki2016}, respectively. Computation time of the Stanford CIM is 1.6ms, which is the time for 1000 round-trips of the 320-m fiber ring cavity. Since the NTT CIM processes a 2000-node problem in 5.0ms, which is the time for 1000 round-trips of the 1-km fiber ring cavity, we can solve up to $\lfloor 2000/N\rfloor$ problems in parallel per the computation time.

The CIM's reliable operation depends on relative phases between the OPO pulses, injection pulses, and measurement local oscillator pulses being kept stable and well-calibrated. Such phase stabilization is imperfect in the experimental setups used in this study, and consequently post-selection procedures have been applied to both the Stanford and NTT CIM experimental data. This is described in detail in Supp.\ Sec.~\ref{sec:supp2}. Computation times have been reported in terms of annealing times; as with the DW2Q, these times exclude the time required to transfer data to and from the CIM.

\ifdefined\isnature \begin{addendum} \else \subsection*{Acknowledgements}\fi
 \ifdefined\isnature \item \fi This research was funded by the Impulsing Paradigm Change through Disruptive Technologies (ImPACT) Program of the Council of Science, Technology and Innovation (Cabinet Office, Government of Japan).  R.H.\ is supported by an IC Postdoctoral Research Fellowship at MIT, administered by ORISE through U.S.\ DOE and ODNI.  P.L.M.\ was partially supported by a Stanford Nano- and Quantum Science and Engineering Postdoctoral Fellowship.  D.V. acknowledges funding from NASA Academic Mission Services, contract no.\ NNA16BD14C.  H.M., E.N., and T.O.\ acknowledge funding from NSF award PHY-1648807.  D.E.\ acknowledges support from the U.S.~ARO through the ISN at MIT (no.~W911NF-18-2-0048) and the SRC-NSF E2CDA program.  The authors acknowledge Salvatore Mandr\`{a} for useful discussions and parallel-tempering simulation results, and Daniel Lidar, Andrew King, and Catherine McGeoch for helpful correspondence.
 \ifdefined\isnature \item[Author Contributions] \else \\\\ \fi  Y.Y., P.L.M.\ and E.R.\ proposed the project.  R.H.\ performed \dwave experiments and data analysis, and prepared the figures. R.H., P.L.M., D.V., and T.I.\ wrote the manuscript.  T.I.\ and P.L.M.\ performed NTT and Stanford CIM experiments, respectively.  D.V.\ helped with \dwave experiments and data analysis.  A.M., C.L., R.L.B., M.M.F., and H.M.\ contributed to building the Stanford CIM.  K.I., T.H., K.E., T.U., R.K., and H.T.\ contributed to building the NTT CIM.  R.H.\ performed simulations of the CIM, adapting code from P.L.M., E.N., and T.O.  E.R., Y.Y., and A.M.\ assisted with preparation of the manuscript.  S.U., S.K., K.K., and D.E.\ assisted with interpretation of the results.
 \ifdefined\isnature \item[Competing Interests] \else \\\\ \fi T.I., H.T., T.H., S.U., Y.Y.\ are inventors on patent JP6429346 awarded in 11/2018 to National Institute of Informatics (NII), Nippon Telegraph and Telephone Corporation (NTT), and Osaka University that covers an Ising model quantum computation device.  
A.M., Y.Y., R.L.B., and S.U.\ are inventors on patent US9830555 awarded in 11/2017 to Stanford University that covers a coherent Ising machine based on a network of optical parametric oscillators.  
S.U., Y.Y., and H.T.\ are inventors on patent US10140580 awarded in 11/2018 to NII and NTT that covers a coherent Ising machine utilizing measurement feedback.  
T.I., K.I., H.T., and T.H.\ are inventors on patent application PCT/JP2018/038994 submitted by NTT that covers a phase checking scheme for the coherent Ising machine.  
T.U.\ and K.E.\ are inventors on patent JP5856083 awarded in 02/2016 to NTT that covers phase-sensitive amplifiers based on periodically poled lithium niobate waveguides.  
P.L.M.\ is an advisor to QC Ware Corp.
 \ifdefined\isnature \item[Correspondence] \else \\\\ \fi Correspondence and requests for materials
should be addressed to R.H.~(e-mail: rhamerly@mit.edu), T.I.~(e-mail: takahiro.inagaki.vn@hco.ntt.co.jp), or P.L.M.~(e-mail: pmcmahon@stanford.edu).
\ifdefined\isnature \end{addendum} \fi


\newpage

\section*{[Supplementary] Experimental investigation of performance differences between Coherent Ising Machines and a quantum annealer}
\renewcommand{\thetable}{S\arabic{table}}
\renewcommand{\thefigure}{S\arabic{figure}}
\renewcommand{\thesection}{S\arabic{section}}
\renewcommand{\thesubsection}{S\arabic{section}.\arabic{subsection}}
\renewcommand{\theequation}{S\arabic{equation}}
\setcounter{section}{0}
\setcounter{figure}{0}
\setcounter{table}{0}
\setcounter{equation}{0}

\section{\dwave embeddings and $J_c$ optimization}
\label{sec:supp1}

Native clique embeddings \cite{Boothby2016} are used for all SK problems, \maxcut problems on graphs with edge density 0.5, and \maxcut problems on varying-density graphs (Figs.~\ref{fig:f2}(c), \ref{fig:f3}(b) and \ref{fig:f4}(a) respectively in main text).  The code to generate the embeddings is available on GitHub \cite{EmbeddingCode}.  Once an embedding is chosen, the embedding parameter $J_c$ (ferromagnetic coupling between qubits in a chain) is tuned to maximize performance.  In no cases does the optimal $J_c$ depend on the annealing time.

\begin{figure}[b!]
\begin{center}
\includegraphics[width=1.00\textwidth]{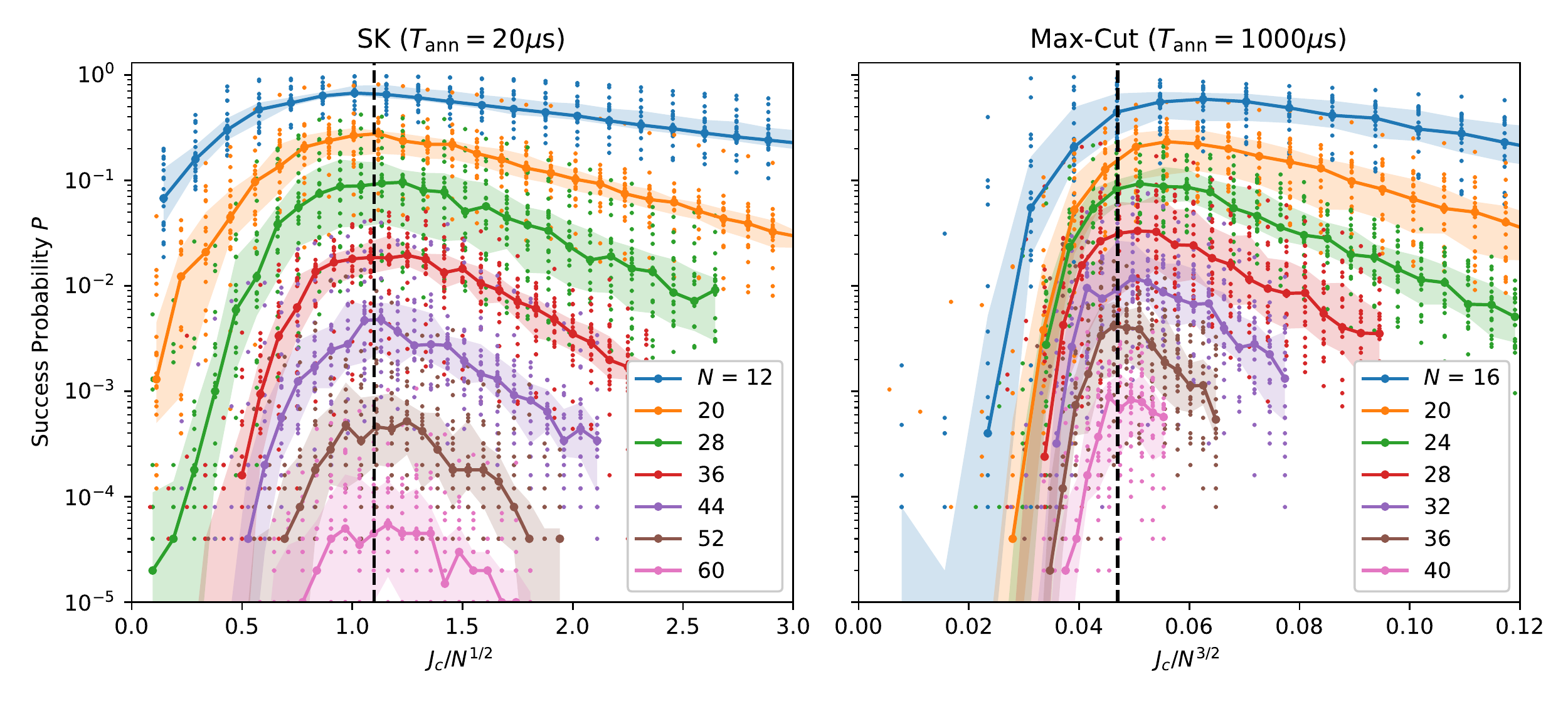}
\caption{\dwave success probability for SK problems and \maxcut problems of edge density 0.5, as a function of problem size $N$ and embedding parameter $J_c$.}
\label{fig:fs1}
\end{center}
\end{figure}

Fig.~\ref{fig:fs1} shows that the optimal $J_c$ scales roughly as $N^{1/2}$ for SK problems and $N^{3/2}$ for \maxcut problems of edge density 0.5.  In particular, the relations $J_c = 1.1 N^{1/2}$ (SK) and $J_c = 0.047 N^{3/2}$ (\maxcut) were used in Figs.~\ref{fig:f2}(c), \ref{fig:f3}(b).

For graphs with variable edge density, it was shown in Fig.~\ref{fig:f4}(b) that the optimal $J_c$ scales as $d$ for fixed $N$, with $J_c = 0.5 d = 9.5x$ for $N = 20$ shown in the figure ($x = d/(N-1)$ is the edge density).  Extrapolating this using the $N^{3/2}$ relation above (which holds for constant $x = \tfrac{1}{2}$), we used $J_c = 9.5(N/20)^{3/2} x$, which is very close to the $J_c = 0.047 N^{3/2}$ used for edge-density 0.5 graphs.  The relation was also tested for $N = 30$ variable edge-density graphs and found to give the optimal $J_c$.

\begin{figure}[t]
\begin{center}
\includegraphics[width=0.6\textwidth]{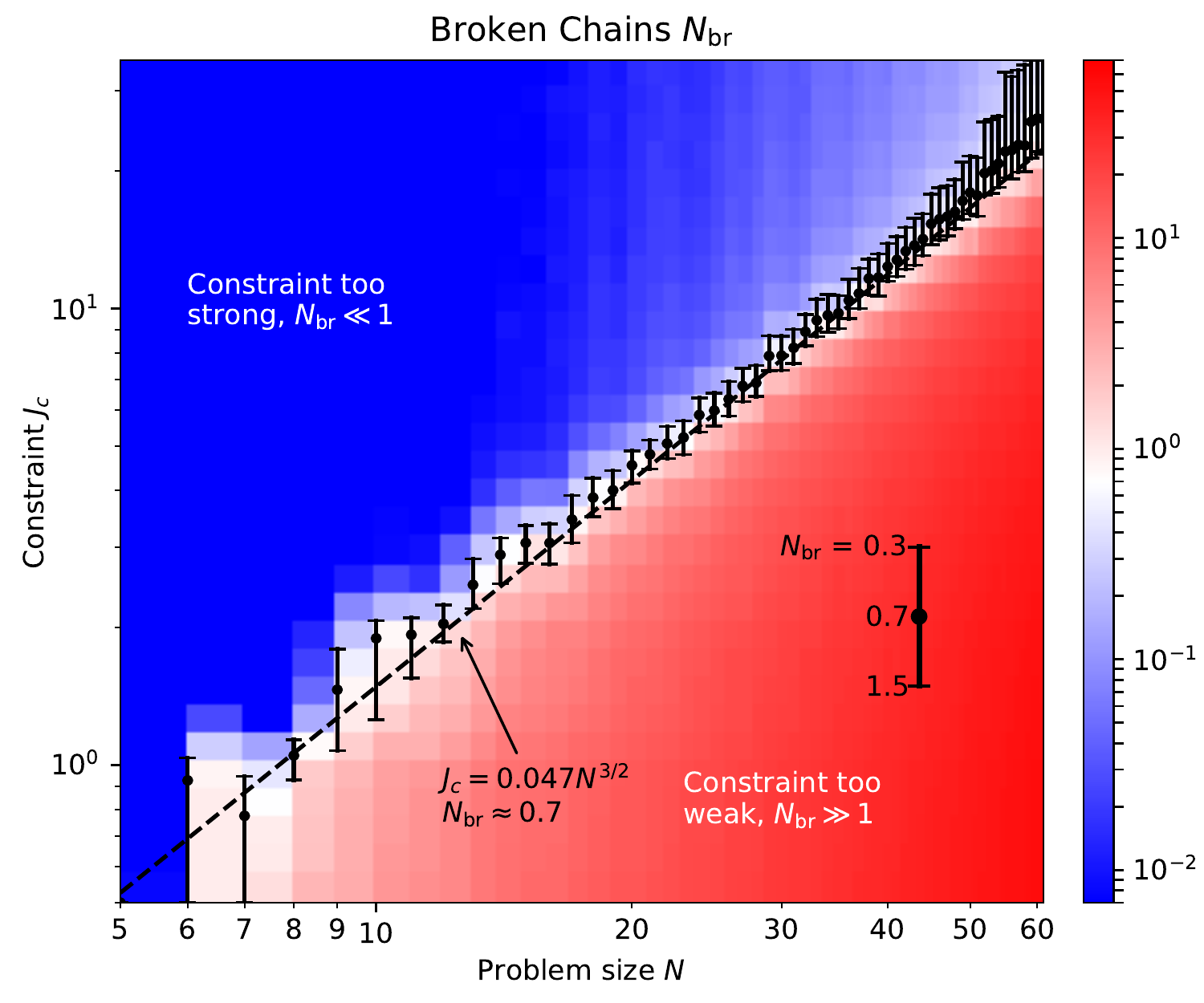}
\caption{\maxcut on edge-density 0.5 graphs.  Broken chains as a function of problem size $N$ and embedding parameter $J_c$}
\label{fig:fs2}
\end{center}
\end{figure}

Fig.~\ref{fig:f3}(a) of the main text suggests that the success probability is maximized when the number of broken chains is $N_{\rm br} \approx 0.7$.  Plotting $N_{\rm br}$ as a function of $N$ and $J_c$ in Fig.~\ref{fig:fs2}, we see that $N_{\rm br} \approx 0.7$ for a narrow range of $J_c$ centered around the line $J_c = 0.047 N^{3/2}$.  For a wide range of $N$, this value of $J_c$ also roughly maximizes the success probability (Fig.~\ref{fig:fs1}).

\begin{figure}[t!]
\begin{center}
\includegraphics[width=1.00\textwidth]{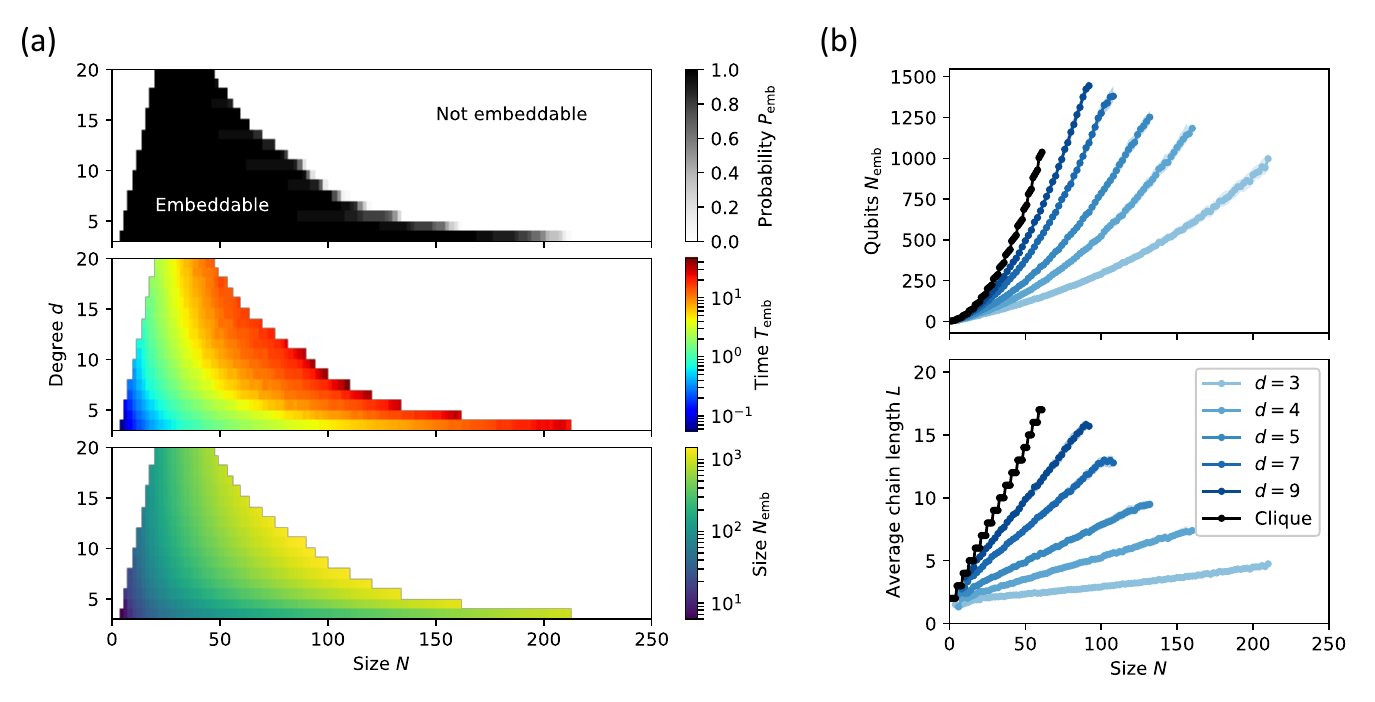}
\caption{Properties of heuristic embeddings for fixed-degree graphs.  ({\bf A}) Probability of finding an embedding using the heuristic, average time required to find an embedding, and number of physical qubits as a function of graph parameters $(N, d)$ for fixed-degree graphs.  ({\bf B}) Number of qubits and average embedding chain length as functions of $N$.}
\label{fig:fs3}
\end{center}
\end{figure}

The fact that dense \maxcut problems are optimally embedded when $N_{\rm br} = O(1)$ is an example of the general principle that $J_c$ must neither be too strong nor too weak for a problem.  If $J_c$ is too small so that $N_{\rm br} \gg 1$, the constraint is not enforced effectively and thus the embedded problem can have a ground state that is different from the logical problem.  Once $N_{\rm br} \lesssim 1$, increasing $J_c$ further will not improve the computation significantly because all of the constraints are already satisfied with high probability.  Rather, it degrades performance because $J_c$ maxes out the physical coupling on the chip so that logical couplings are scaled down as $J_c^{-1}$, which will correspondingly reduce the spectral gap of the (physical) Hamiltonian, and can also cause problems due to the finite bit precision and hardware imperfections of the \dwave system.

For the sparse graphs, embeddings are found using the heuristic of Cai et al.~\cite{Cai2014}, which is available as part of the \dwave API toolkit.  For each sparse graph instance, we attempt to generate 10 embeddings using the heuristic with a time-out of 60 seconds.  The probability of finding an embedding is shown in Fig.~\ref{fig:fs3}(a) (the $d = 3$ case is in agreement with \cite[Fig.~7]{Cai2014}).  The time required to find an embedding (on average) and the number of physical qubits $N_{\rm emb}$ are also plotted in Fig.~\ref{fig:fs3}(a).

\begin{figure}[t!]
\begin{center}
\includegraphics[width=1.00\textwidth]{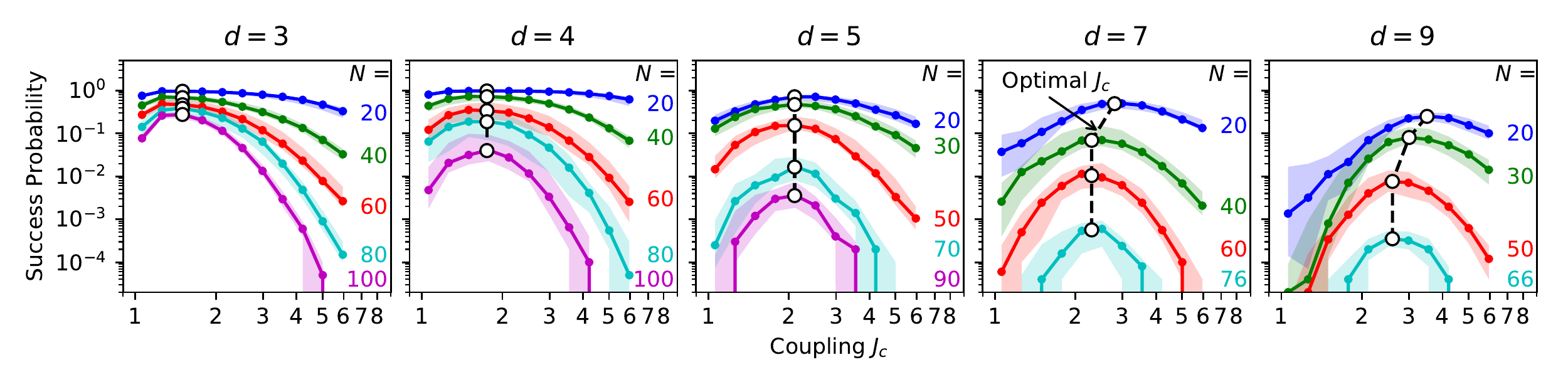}
\caption{Choice of optimal coupling for sparse graphs using the heuristic embedding.}
\label{fig:fs4}
\end{center}
\end{figure}

Fig.~\ref{fig:fs3}(b) shows the number of physical qubits for graphs of degree $d = 3, 4, 5, 7, 9$ embedded using the heuristic, as well as the average chain length $L = N_{\rm emb}/N$.  This is compared against the clique embeddings described above.

Because the heuristic embeddings differ markedly from clique embeddings, we do not use the formula $J_c = 9.5(N/20)^{3/2} x$ derived above.  Rather, the optimal $J_c$ is found by hand, running the quantum annealer for a range of $N$, $d$ and $J_c$ (Fig.~\ref{fig:fs4}).  We find that the optimal $J_c$ is independent of $N$ for sufficiently large $N$, while it increases slightly for small $N$ for $d = 7, 9$.  We interpolate using the curves of Fig.~\ref{fig:fs4} to find the embedding parameter used in the main text (Fig.~\ref{fig:f3}(c)).

\section{CIM data and post-selection}
\label{sec:supp2}

The CIM is based on an OPO network, which is sensitive to optical phase fluctuations.  During the course of operation, the phase of the injection beam will drift.  This drift is slow compared to experimental timescales, but can become large if a calculation is run thousands of times.

To filter out out-of-phase computations (which always lead to the wrong answer), each CIM includes a phase-checking mechanism, albeit somewhat different for the NTT and the Stanford CIMs. We summarize both here.

\begin{figure}[tb]
\begin{center}
\includegraphics[width=1.00\textwidth]{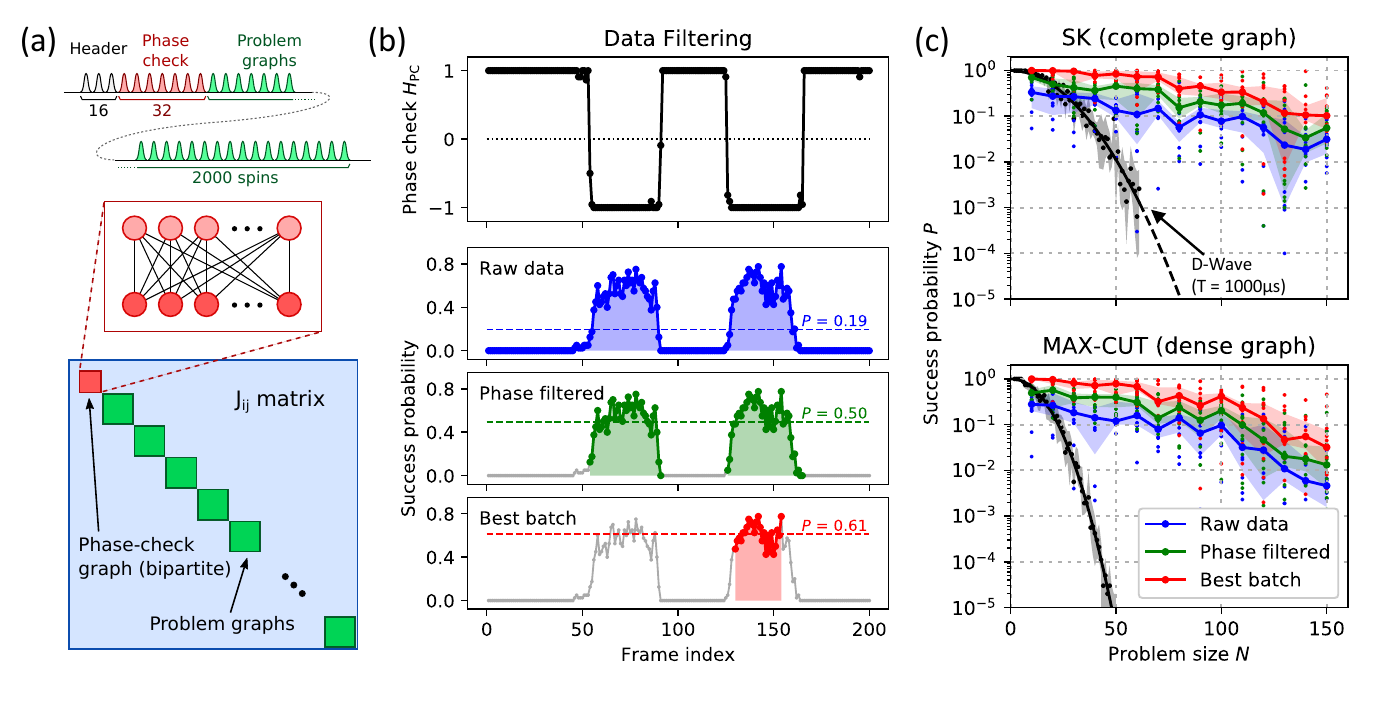}
\caption{Data filtering and post-selection in NTT CIM.  ({\bf A}) Partitioning of NTT CIM spins into a 16-spin header, a 32-spin phase-check graph, and 2,000 spins for problem graphs, and the resulting $J_{ij}$ matrix.  ({\bf B}) Phase-check Hamiltonian $H_{\rm PC}$ as a function of time (frame index), and three post-selection techniques for inferring the success probability.  ({\bf C}) NTT CIM success probability for SK and $x = 0.5$ \maxcut problems as a function of post-selection method.}
\label{fig:fs5}
\end{center}
\end{figure}

In the NTT system, phase stability and calibration is implemented with a phase-check graph: the 2,048 spins in the CIM are partitioned into a 16-spin (unused) header, a 32-spin bipartite graph for phase checking, and a ``frame'' of 2,000 spins for the desired problem.  Since $N \ll 2000$ for the problems in this paper, we can solve up to $\lfloor 2000/N \rfloor \approx 2000/N$ problems in parallel per frame.  The coupling matrix $J_{ij}$ has a block-diagonal structure (Fig.~\ref{fig:fs5}(a)).

The couplings of the bipartite graph for phase-check are randomly set to $+1$ or $-1$ and the value of the phase-check Hamiltonian $H_{\rm PC} = \frac{1}{2}\sum_{ij} J_{ij}\sigma_i\sigma_j$ is computed after each run.  If the optical phase is incorrect, we find $H_{\rm PC} > 0$ because the system couplings are reversed and the machine is trying to minimize $-H_{\rm PC}$.  The top plot of Fig.~\ref{fig:fs5}(b) shows the phase-check $H_{\rm PC}$ value (normalized to the maximum) as a function of time.  $H_{\rm PC}$ drops sharply to a negative value when the CIM is in phase, making it a good proxy for the CIM phase.

In the bottom plots of Fig.~\ref{fig:fs5}(b), three data-filtering techniques are shown.  Here we plot the free-running success probability (fraction of instances per frame in the ground state) for an $N = 50$ problem (40 trials running in parallel per frame).  Averaging over all frames requires no post-processing, but gives a low success probability because we are including many trials when the machine is out of phase.  Filtering on the phase-check graph (green curve) does significantly better; however, we are still averaging over the edges of the phase-check region where the system is only marginally in phase.  Still better success probabilities can be found by looking for the best batch of 1,000 consecutive trials (25 consecutive frames) in the series (red curve).  This generally corresponds to the the CIM working in its best condition: when the feedback signal is well in phase.  This is the success probability we could expect from a well-engineered CIM where the optical phase, pump power, and other optical degrees of freedom have been sufficiently stabilized.

We compare the three post-selection methods in Fig.~\ref{fig:fs5}(c) to show that our post-selection techniques give only a constant improvement in success probability, and this constant is never more than an order of magnitude.  Thus, we can safely conclude that the CIM's performance advantage does not arise from cherry-picking good samples from the data.  The ``best batch'' method (red curves in Fig.~\ref{fig:fs5}) is used to process all CIM data reported in the main text.

The data collected from the Stanford CIM was also post-processed to select only the runs on the machine for which the optical setup was optimally stable. However, the procedure for post-selection was slightly different to that used for the data from the NTT CIM.  In the case of the Stanford CIM, a recording of the homodyne measurement of the output pulses immediately before a run began was stored. During this recording phase, constant-amplitude pulses were injected into the cavity. If the entire system is phase-stable, then the recorded homodyne measurement results should not show large fluctuations from pulse to pulse. Furthermore, the particular value of the phase of the injected light is also relevant (not just that it is ideally constant), since the computation mechanism relies on interference of injected pulses with pulses in the cavity, and how much interference is obtained is partially determined by the phase of the injection pulses. We therefore post-selected not only for stability, but also for a particular mean value of the homodyne measurement results, which was determined on an instance-by-instance basis. The net effect of this post-selection procedure is to produce success probabilities that represent the probabilities one would obtain if the CIM was always phase-stable whenever a computation was run, and the phase was correctly calibrated for each problem instance.

The post-selected success probabilities were only on average $5\times$ higher than the success probabilities obtained when no post-selection was applied. 
This implies that even if one is pessimistic about the prospects of improvement to the optical phase stabilization of the CIM, and one assumes that the most stable the machine will ever be is as it was during the experiments reported in this paper, then at worst one should divide the success probabilities for the Stanford CIM reported in this paper by $5\times$. This gives the estimate for the expected success probabilities for a machine that has the same fundamental operating principle as the currently implemented CIM at Stanford, as well as the same experimental imperfections (including phase noise) that the current setup has.

\begin{figure}[b!]
\begin{center}
\includegraphics[width=1.00\textwidth]{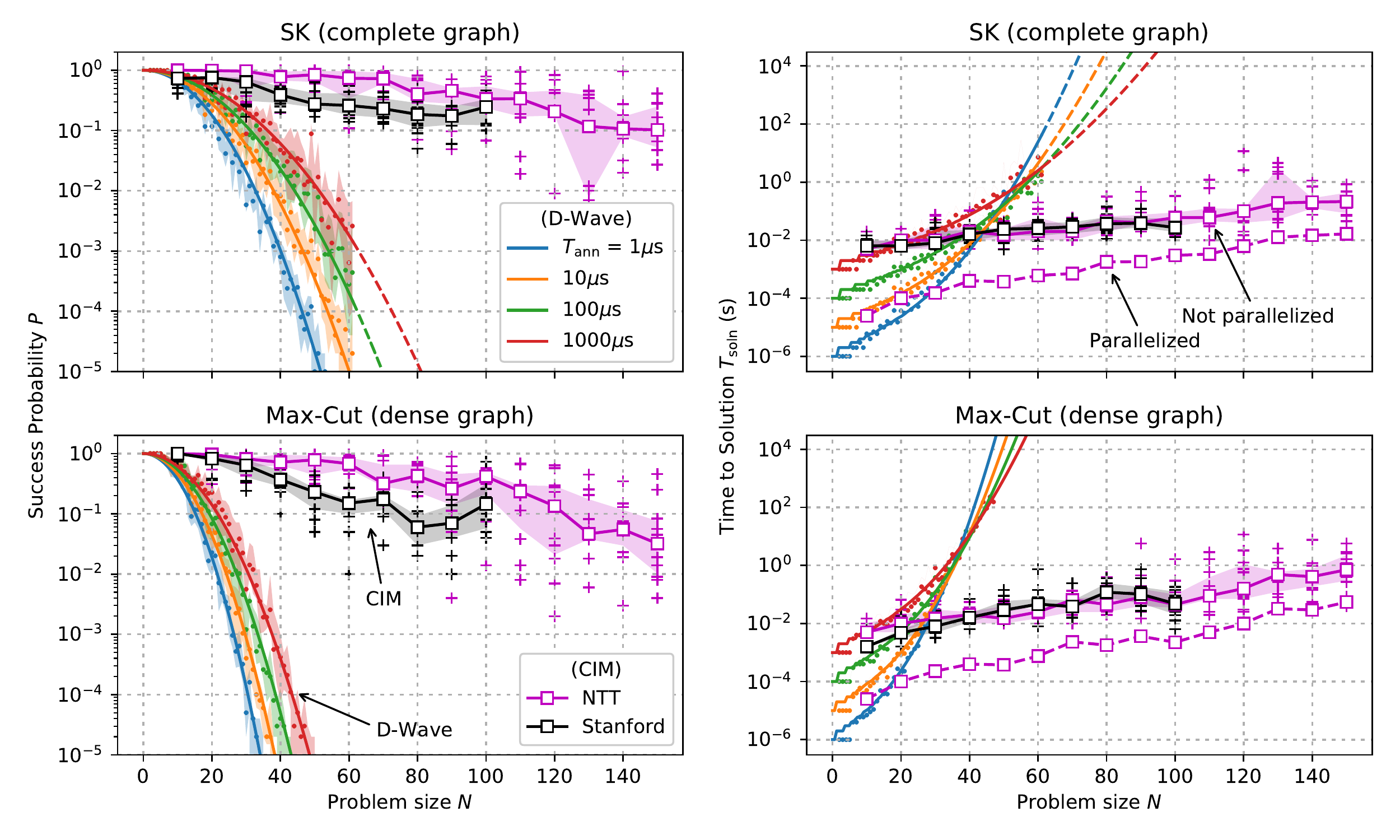}
\caption{Comparison of Stanford and NTT CIM performance for SK and dense \maxcut problems.  Annealing time is 1000 round trips.  D-Wave data for $T_{\rm ann} = 1$, $10$, $100$, and $1000\upmu{\rm s}$ are also plotted.}
\label{fig:fs6}
\end{center}
\end{figure}

The CIMs at Stanford and NTT were run on the same (randomly-chosen) Ising problems for $N \leq 100$ \maxcut (edge density $x = 0.5$) and SK (fully connected).  The average success probabilities of the two machines agree to within a factor of 5 (Fig.~\ref{fig:fs6}).

In order to compare the solution time $T_{\rm soln}$ with \dwave, we need the physical annealing time for the CIM.  A strict minimum for the annealing time is given by the product of the time between pulses (equal to $1/f$ where $f$ is the pump repetition frequency), the size of the problem $N$, and the number of round trips per run $R$:
\beq
	T_{\rm ann}^{\rm (min)} = \frac{N R}{f} \label{eq:tann}
\eeq
This is the effective annealing time if perfect parallelization is achieved and all spins are used for logic (i.e.\ a negligible fraction of phase-check and dummy spins).  Both Stanford and NTT CIMs use $R = 1000$ round trips.

However, the annealing time is generally longer than $T_{\rm ann}^{\rm (min)}$ because dummy spins are added to the cavity to compensate for the delays due to the DAC / ADC electronics in the feedback circuit and to give the FPGA more time to finish the coupling computation.  This increases the cavity round-trip time and thus the annealing time.

In the NTT CIM, we used 5056 pulses in a 1-km fiber ring cavity as: 16-spin (header), 32-spin (phase check), 2000-spin (solve problem), 100-spin (blank), 2808-spin (free running in FPGA calculation time), 100-spin (blank).  The pump repetition rate is 1 GHz and the round-trip time is 5$\upmu$s.  As only 2000 of 5056 pulses are used, even if perfect parallelism is employed, the annealing time is approximately $2.5\times$ longer than Eq.~(\ref{eq:tann}), or $T_{\rm ann} = (2.5N)\upmu{\rm s}$, where $N$ is the problem size.  Fig.~\ref{fig:fs6} plots the NTT CIM time-to-solution both with and without parallelism, to enable a fair comparison with the \dwave annealer (we did not attempt to parallelize \dwave to run multiple problems per anneal).

In the Stanford CIM, which did not employ parallelism due to its smaller number of spins, the annealing time is $T_{\rm ann} = 1.6\,{\rm ms}$ for all problems.  The Stanford CIM \cite{McMahon2016} features a 320-m fiber ring cavity that contains 160 optical pulses (repetition rate 100 MHz), of which up to 100 can be used to encode Ising problems.  The data in Fig.~\ref{fig:f2}(c) come from the Stanford CIM, where the above annealing time combined with the formula $T_{\rm soln} = T_{\rm ann} \lceil \log(0.01)/\log(1 - P)\rceil$ is used to calculate the time to solution.

\section{C-SDE Simulations of CIM}
\label{sec:supp3}

\begin{figure}[b!]
\begin{center}
\includegraphics[width=0.5\textwidth]{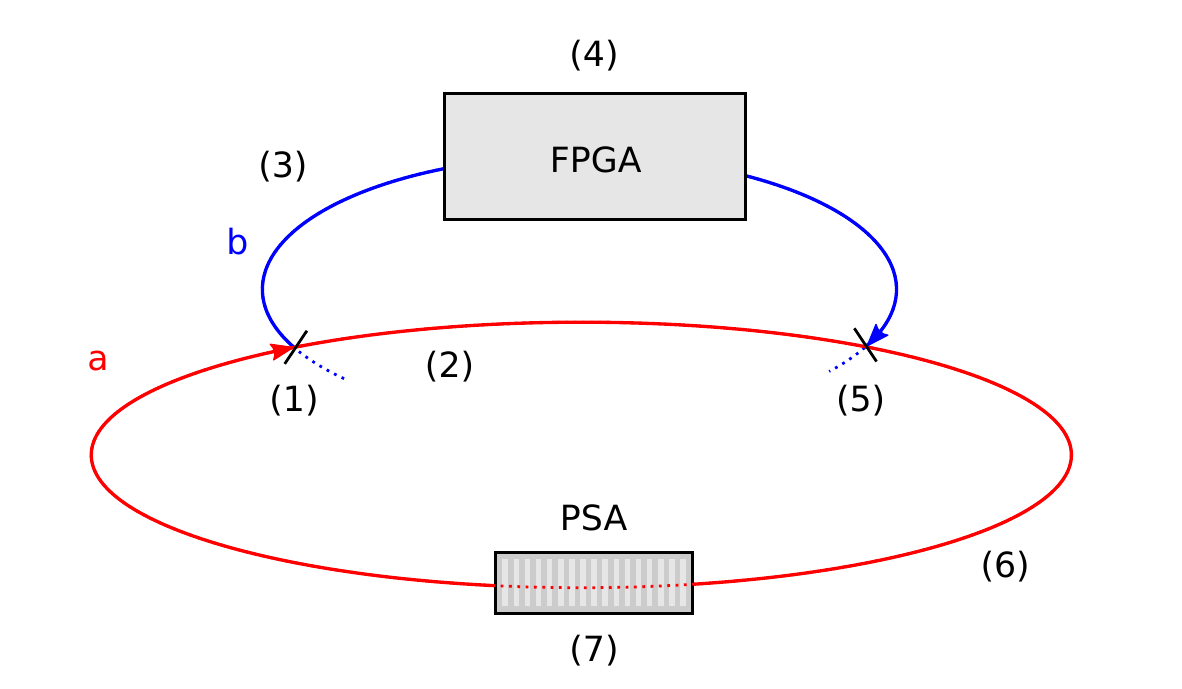}
\caption{Abstract schematic of measurement-feedback CIM.  State variables are fields $a$, $b$.  Each process (1)--(7) is described in Table \ref{tab:ts1}.}
\label{fig:fs7}
\end{center}
\end{figure}

The CIM is a time-multiplexed synchronously-pumped OPO with measurement feedback coupling (Fig.~\ref{fig:fs7}).  It consists of a main loop (red) with a delay line for measurement and feedback (blue).  Because the OPO is weakly coupled, we can treat this system using truncated-Wigner theory \cite{Kinsler1991}, which reduces the quantum dynamics to a set of c-number Langevin equations (c-SDEs).  For OPOs with low single-pass gain at threshold, continuous-time stochastic differential equations can be employed \cite{Wang2013}.  Since the round-trip gain of our system is high, a discrete-time model is needed, where the evolution of a single round trip (represented as a discrete-time c-SDE) consists of a series of seven discrete steps, each with its appropriate truncated-Wigner description (Table \ref{tab:ts1}).

\begin{table}[t!]
\begin{center}
\begin{tabular}{c|cc}
\hline\hline
Step & Description & Truncated-Wigner Model \\ \hline
1 & Beamsplitter & $a_i \cos(\theta_m) + w^{(1)}_i \sin(\theta_m) \rightarrow a_i$ \\ 
& & $a_i \sin(\theta_m) - w^{(1)}_i \cos(\theta_m) \rightarrow b_i$ \\ \hline
2 & Loss & $a_i \cos(\theta_{L1}) + w^{(2)}_i \sin(\theta_{L1}) \rightarrow a_i$ \\ \hline
3 & Loss & $b_i \cos(\theta_{L2}) + w^{(3)}_i \sin(\theta_{L2}) \rightarrow b_i$ \\ \hline
4 & Detection & $b_i \rightarrow x_i$ \\
& FPGA & $\sum_j J_{ij} x_j \rightarrow y_i$ \\
& Modulation & $C(F(t) y_i; y_{\rm max}) + w^{(4)}_i \rightarrow b_i$ \\ \hline
5 & Beamsplitter & $a_i \cos(\theta_f) + b_i \sin(\theta_f) \rightarrow a_i$ \\ \hline
6 & Loss & $a_i \cos(\theta_{L3}) + w^{(5)}_i \sin(\theta_{L3}) \rightarrow a_i$ \\ \hline
7 & PSA Gain & $p + w^{(6)}_i \rightarrow p_i$ \\
& & $\epsilon L\sqrt{p_i^2 + a_i^2/2} \rightarrow B_i$ \\
& & $e^{B_i} \bigl(1 + \tfrac{1}{2}(e^{2B_i}-1)(1 - (1 + a_i^2/2p_i^2)^{-1/2})\bigr) a_i \rightarrow a_i$ \\
\hline\hline
\end{tabular}
\caption{Seven steps in a single round trip for the measurement-feedback CIM, and the appropriate truncated-Wigner description.  Constants are chosen to match the Stanford CIM: $\epsilon L = 3.6\times 10^{-4}$, $p = 2.8 \times 10^{3}$, $\sin(\theta_m) = \sin(\theta_f) = \sqrt{0.1}$, $\sin(\theta_{L1}) = \sqrt{0.6}$, $\sin(\theta_{L2}) = \sqrt{0.5}$, $\sin(\theta_{L3}) = \sqrt{0.6}$.}
\label{tab:ts1}
\end{center}
\end{table}

Steps 1 and 5 are standard beamsplitters, whose input/output equations match those in classical optics.  Steps 2, 3 and 6, which represent loss in the fiber loop and injection channel, can be modeled as beamsplitters with vacuum inputs.  Homodyne detection converts the real part of $b_i$ to a classical signal, discarding the imaginary part (only the real parts of optical signals $a_i$, $b_i$ are treated in this model).  The resulting classical signal is processed in the FPGA (step 4).  The FPGA result is imprinted onto an optical field using a modulator, adding the vacuum fluctuations of the injected field.  The modulation signal is clamped (function $C(z; z_0) \equiv \max(\min(z, z_0), -z_0)$) by the DAC maximum voltage (parameter $y_{\rm max}$ above).  Step 7 is the $\chi^{(2)}$ phase-sensitive amplifier (PSA) gain.  The formula is derived by solving the nonlinear field equations \cite[Sec.~2.2]{Boyd2003} in a $\chi^{(2)}$ medium \cite[Eq.~(8)]{Hamerly2016}.  All input vacuum fields are normally distributed random variables: $w^{(m)}_i \sim N(0, \tfrac{1}{2})$.  We note that our model bears resemblance to mean-field annealing approaches to the Ising problem \cite{King2018}.

In this paper, the constants are chosen to match the experimental parameters of the Stanford CIM.  The model is sensitive to the measurement and feedback couplings $(\theta_m, \theta_f)$, but less sensitive to the overall loss, which simply increases the amount of quantum noise in the system by a small amount.  

The Stanford CIM employs an ``injection turn-on'' scheme.  We start with the feedback turned off and pump the OPO to slightly below threshold.  Then the feedback term is slowly increased, lowering the effective threshold of the coupled-OPO system \cite{McMahon2016}.  This is opposite to the ``pump turn-on'' technique used in the NTT CIM and optical-feedback systems \cite{Inagaki2016, Marandi2014, Takata2016}, where the coupling (and therefore threshold) stays fixed and the pump is increased.  But the fundamental dynamics (bifurcation from squeezed vacuum driven by quantum noise) is the same, and we expect similar computational performance for both machines.  The key degree of freedom is the {\it pump schedule} $F(t)$.  For simplicity, we use a linear ramp
\beq
	F(t) = F_{\rm max} \frac{t}{T_{\rm ann}}
\eeq
which increases from zero to $F_{\rm max}$ over $T_{\rm ann}$ round trips (the runtime, or ``annealing time'', of the CIM, where $T_{\rm ann} = 1000$ in the experiments in this paper.)

\begin{figure}[p]
\begin{center}
\includegraphics[width=1.0\textwidth]{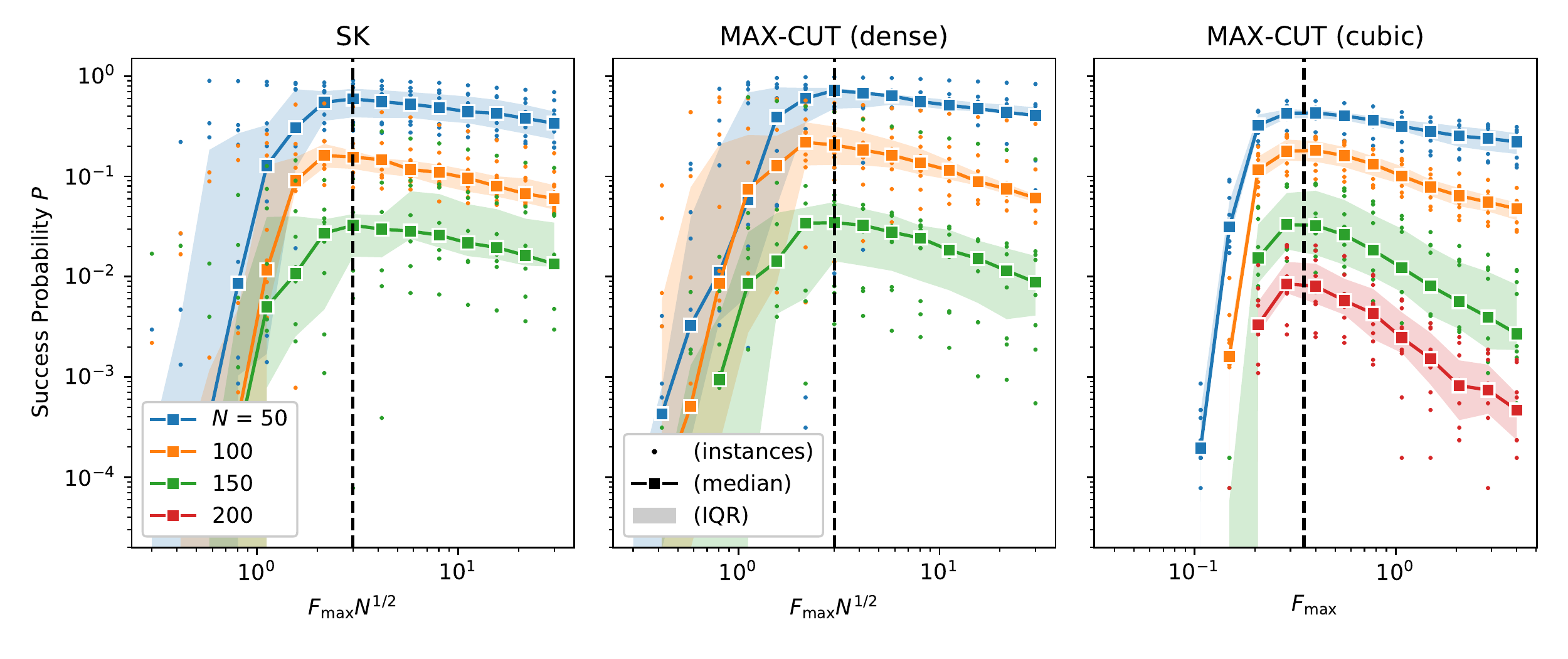}
\caption{Simulated CIM success probability as a function of $F_{\rm max}$ for the SK, dense \maxcut, and cubic \maxcut problems from this paper.}
\label{fig:fs8}
\end{center}
\end{figure}

\begin{figure}[p]
\begin{center}
\includegraphics[width=1.0\textwidth]{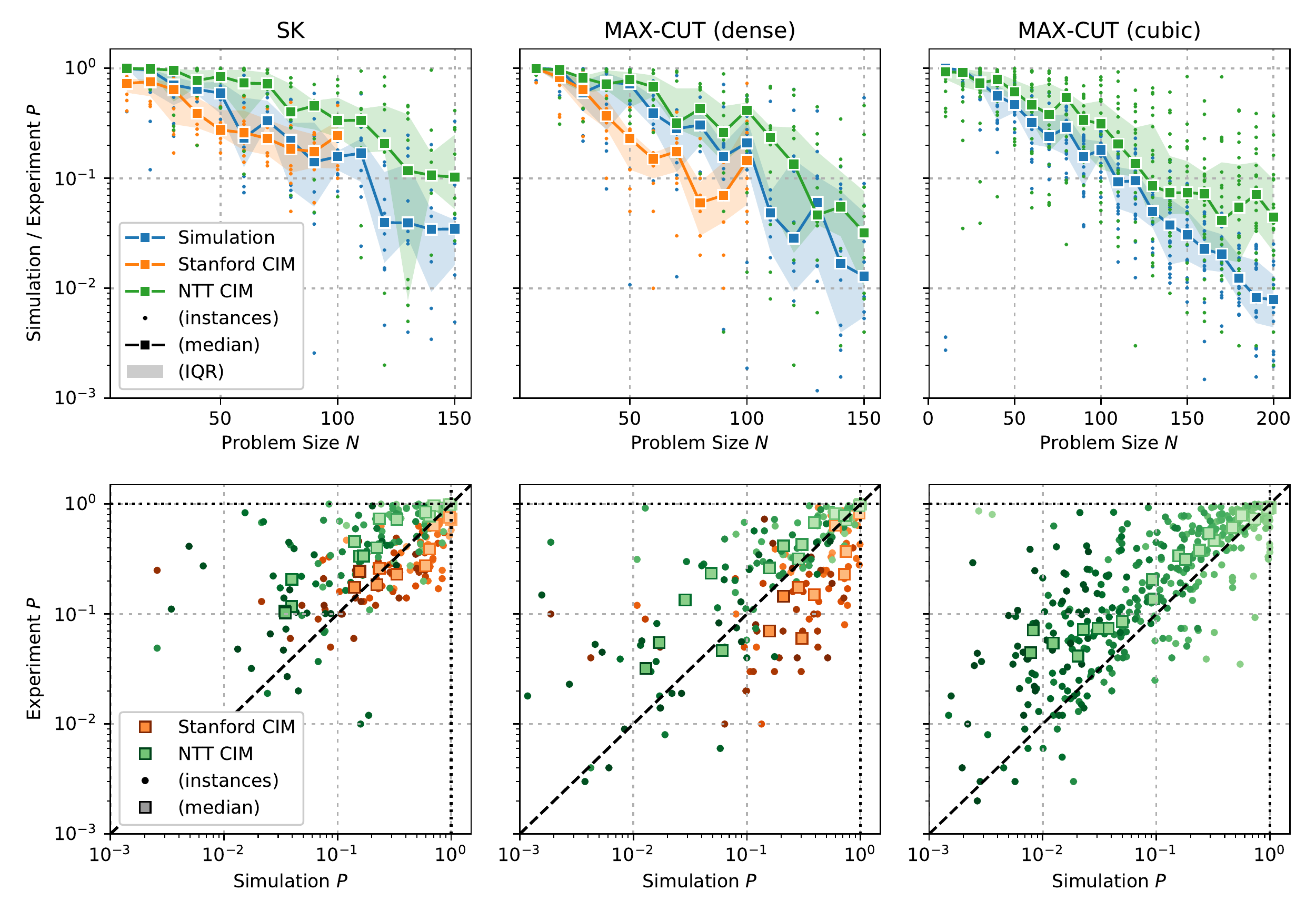}
\caption{Comparison of c-SDE simulations with experimental CIM data.  Top: success probability as a function of problem size.  Bottom: correlation plots between the c-SDE simulated CIM (x-axis) and experimental data (y-axis).}
\label{fig:fs9}
\end{center}
\end{figure}

The free parameter $F_{\rm max}$ sets the scale of the feedback strength, and is tuned to maximize the success probability.  Intuitively, one wants the feedback term $y_i$ to be comparable to the circulating field $a_i$, as a small feedback term will not effectively couple the OPOs but a very large term will lead to spurious behavior that no longer maps onto the Ising problem \cite{Wang2013}.  Since the injected field is proportional to $F(t) \sum_j J_{ij} a_j$, and since in random non-structured problems, the $a_j$ are expected to be random, it is reasonable to assume that:
\beq
	F_{\rm max} \propto \Bigl(\frac{1}{N} \sum_{ij} (J_{ij})^2\Bigr)^{-1/2} \label{eq:ft}
\eeq
For SK and dense \maxcut problems, Eq.~(\ref{eq:ft}) predicts $F_{\rm max} \propto N^{-1/2}$, while for sparse problems, $F_{\rm max}$ should be a constant.  This prediction is confirmed numerically in Fig.~\ref{fig:fs8}.  The success probability depends on both $N$ and $F_{\rm max}$, and the peak is always located at $F_{\rm max}N^{1/2} = \mbox{const}$ for SK and dense \maxcut, and $F_{\rm max} = \mbox{const}$ for sparse \maxcut.  The optimal $F_{\rm max}$ is roughly:
\beq
	F_{\rm max} = \begin{cases}
		3.0 N^{-1/2} & \mbox{(SK)} \\
		3.0 N^{-1/2} & \mbox{(Dense \maxcut)} \\
		0.35 & \mbox{(Cubic \maxcut)} \end{cases} \label{eq:fopt}
\eeq
Using the optimal $F_{\rm max}$ in Eq.~(\ref{eq:fopt}), we simulate the CIM on all of the problems presented in the paper.  Fig.~\ref{fig:fs9} shows the result.  Strictly speaking, the model is only applicable to the Stanford CIM, but both machines give similar performance that is roughly matches the c-SDE simulations.

\begin{figure}[b!]
\begin{center}
\includegraphics[width=1.0\textwidth]{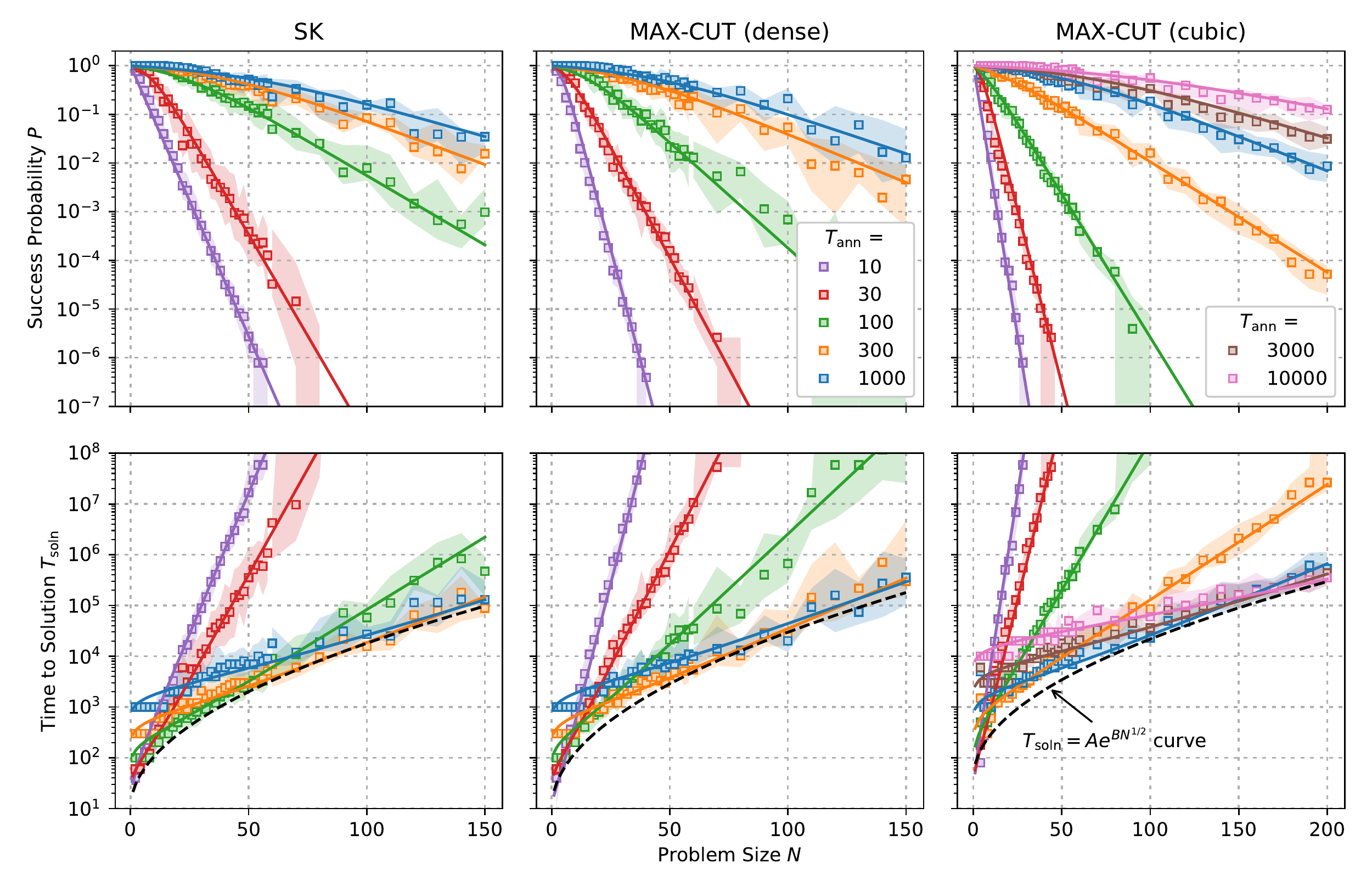}
\caption{Simulated CIM success probability and time to solution (in round trips) for SK and \maxcut problems.  Squares are medians.  Shaded region is IQR.  Solid lines are fits to Eq.~(\ref{eq:fopt}).}
\label{fig:fs10}
\end{center}
\end{figure}

C-SDE simulations are run to assess the effect of the annealing time and to determine the optimal-annealing-time scaling of the CIM time to solution.  Fig.~\ref{fig:fs10} plots the success probability and time to solution (normalized to the round-trip time) for annealing times ranging from $T_{\rm ann} = 10$ round trips up to $T_{\rm ann} = 1000$.  We see clear exponential behavior in the asymptotic limit, especially when $T_{\rm ann}$ is small.  The plots fit reasonably well to a logistic curve intersecting the origin:
\beq
	P(N) = \frac{\alpha}{(\alpha - 1) + e^{\beta N}} \stackrel{N\rightarrow\infty}{\longrightarrow} \alpha e^{-\beta N}
\eeq
Likewise, the time-to-solution curves are rising exponentials in the large-$N$ limit.  When plotted on a logarithmic scale, the intercept of the curves increases with $T_{\rm ann}$, while the slope decreases.  This makes clear that, as in quantum annealing, there is a tradeoff between success probability and annealing time \cite{Ronnow2014}.  The optimal time to solution is given by the lower envelope of these curves.  In quantum annealing on glassy chimera-graph problems, an empirical scaling of $T_{\rm soln} \sim \exp(O(N^{1/2}))$ has been reported \cite{Ronnow2014,Albash2018,Mandra2018}.  Curves of the form $A e^{B N^{1/2}}$ are plotted in Fig.~\ref{fig:fs10} for reference.  The rough fit suggests, but is not conclusive proof of, a similar time-to-solution scaling for coherent Ising machines.

\section{Optimal Anneal-Time Analysis}
\label{sec:supp4}

To obtain the best performance of the \dwave annealer under a fixed anneal schedule, we optimize $T_{\rm soln}$ with respect to the annealing time.  For a fixed $T_{\rm ann}$ we find the square-exponential relation $P = e^{-(N/N_0)^2}$ for SK and dense \maxcut problems.  Cubic \maxcut problems also fit this curve, especially for short anneals.  In the range $T_{\rm ann} \in [1, 2000]\upmu{\rm s}$ of admissible annealing times, we find $N_0 \approx \alpha + \beta \log_{10}(T_{\rm ann}/\upmu{\rm s})$, where $\alpha$ and $\beta$ are problem-dependent constants (Table~\ref{tab:ts2}).

\begin{figure}[t!]
\begin{center}
\includegraphics[width=1.00\textwidth]{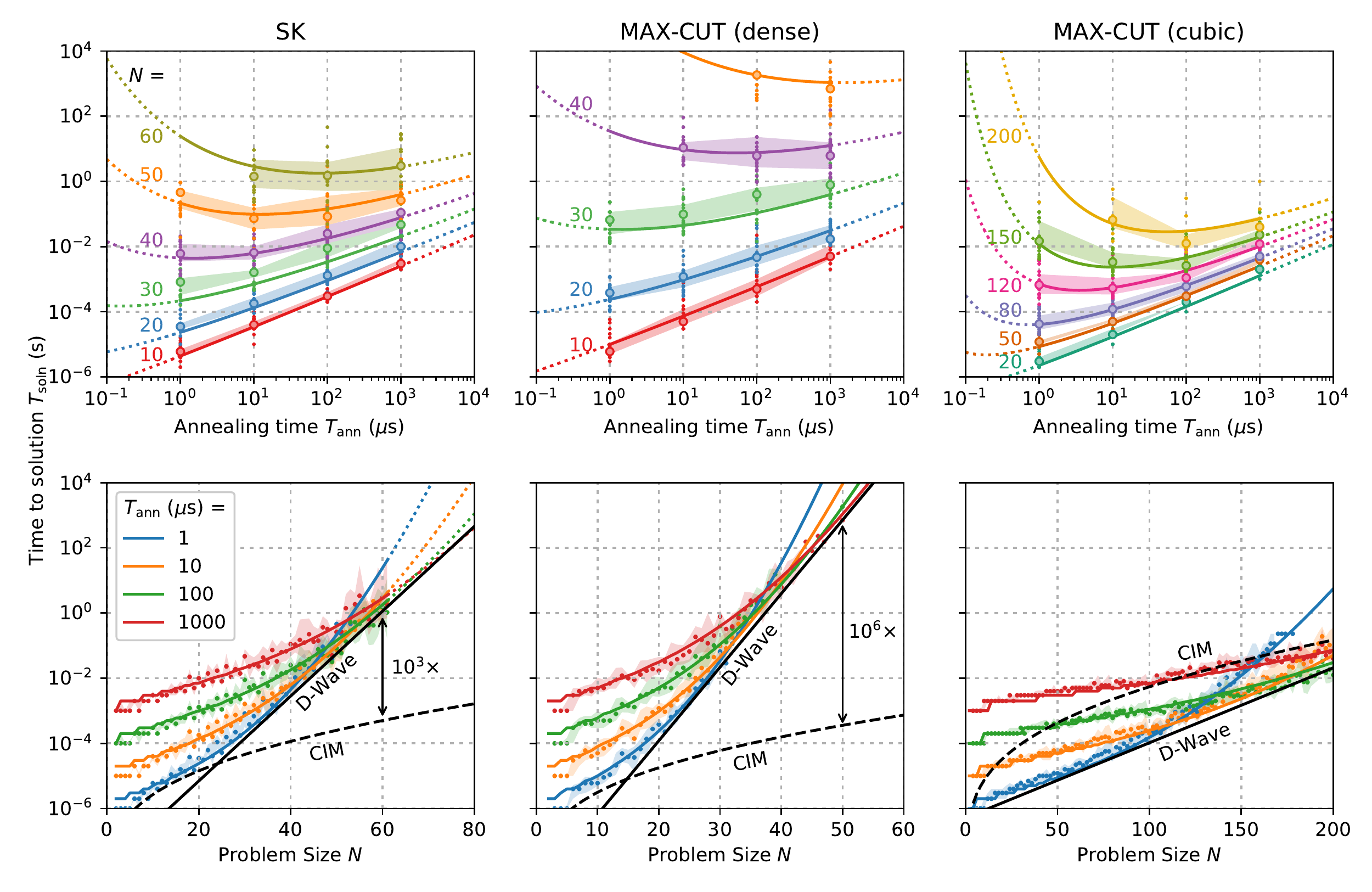}
\caption{Time-to-solution analysis for \dwave at optimal annealing time.  Top: \dwave time to solution $T_{\rm soln}$ as a function of the annealing time for fixed problem sizes, illustrating the optimal anneal time.  Bottom: $T_{\rm soln}$ as a function of problem size, with optimal anneal-time curve approximated as a line.  CIM time to solution at optimal anneal time (from Fig.~\ref{fig:fs10}) plotted for comparison (NTT CIM with parallelization, round-trip time $(2.5 N){\rm ns}$).}
\label{fig:fs11}
\end{center}
\end{figure}

\begin{table}[tb]
\begin{center}
\begin{tabular}{c|ccc}
\hline\hline
& SK & \maxcut (dense) & \maxcut(cubic) \\ \hline
$\alpha$ & 15.24 & 10.05 & 53.45 \\
$\beta$ & 2.81 & 1.39 & 22.15 \\ \hline\hline
\end{tabular}
\caption{Problem-dependent constants $\alpha, \beta$ used in the relation $N_0 = \alpha + \beta\log_{10}(T/\upmu{\rm s})$ for the success-probability exponential $P = e^{-(N/N_0)^2}$}
\label{tab:ts2}
\end{center}
\end{table}

The top graphs in Fig.~\ref{fig:fs11} plot the dependence of $T_{\rm soln}$ on $T_{\rm ann}$ for fixed $N$, allowing one to visualize the optimal annealing time for each problem size.  The aforementioned fit agrees reasonably with the data for most problem sizes, although we make no claims about its validity outside the range of annealing times tested.  

The lower plots in Fig.~\ref{fig:fs11} show the \dwave time to solution in terms of problem size.  The lower envelope of the fixed-$T_{\rm ann}$ curves, approximated as a line ($T_{\rm soln} = A e^{B N}$), gives the optimal time to solution for the DW2Q.  For comparison, the optimal CIM time-to-solution obtained in Fig.~\ref{fig:fs10} is also plotted.  The CIM round-trip time used is the value for the NTT CIM accounting for parallelization: $(2.5N){\rm ns}$; see Sec.~\ref{sec:supp2}.

Since the optimal annealing time lies in the experimentally accessible regime $[1, 2000]\upmu{\rm s}$ for only a limited range of problem sizes ($N \in [40, 60]$ for SK, $[30, 50]$ for dense \maxcut), it is difficult to estimate the precise shape of the lower envelope by looking at Fig.~\ref{fig:fs11}.  While the data is consistent with an exponential, it is also consistent with many other curves, so we caution against naively extrapolating these curves.  Nevertheless, at optimal annealing time, the CIM is substantially faster ($\geq 10^3\times$ for SK, $\geq 10^6\times$ for dense \maxcut) at the upper end of experimentally measured problem sizes, while \dwave has a performance advantage of 10--100$\times$ for cubic \maxcut, although this advantage narrows with larger problem sizes.

\section{Performance of Parallel Tempering}
\label{sec:supp5}

Parallel tempering is a state-of-the-art classical optimization technique that has been shown to perform well on a variety of Ising problems \cite{Mandra2018,Mandra2017,Mandra2016}. Here, we include results provided by Salvatore Mandr\`a, which made use of the implementation of parallel tempering in the NASA/TAMU Unified Framework for Optimization (UFO). The comparison shows respectable performance of NTT's parallel CIM compared with PT@UFO. We see that NTT's parallel CIM comes close to the performance of PT@UFO for the SK problem instances in the size range considered, and is also close on the \maxcut problem up through the middle range of problem sizes considered, but diverges for larger problem sizes.

\begin{figure}[h]
\begin{center}
\includegraphics[width=0.8\textwidth]{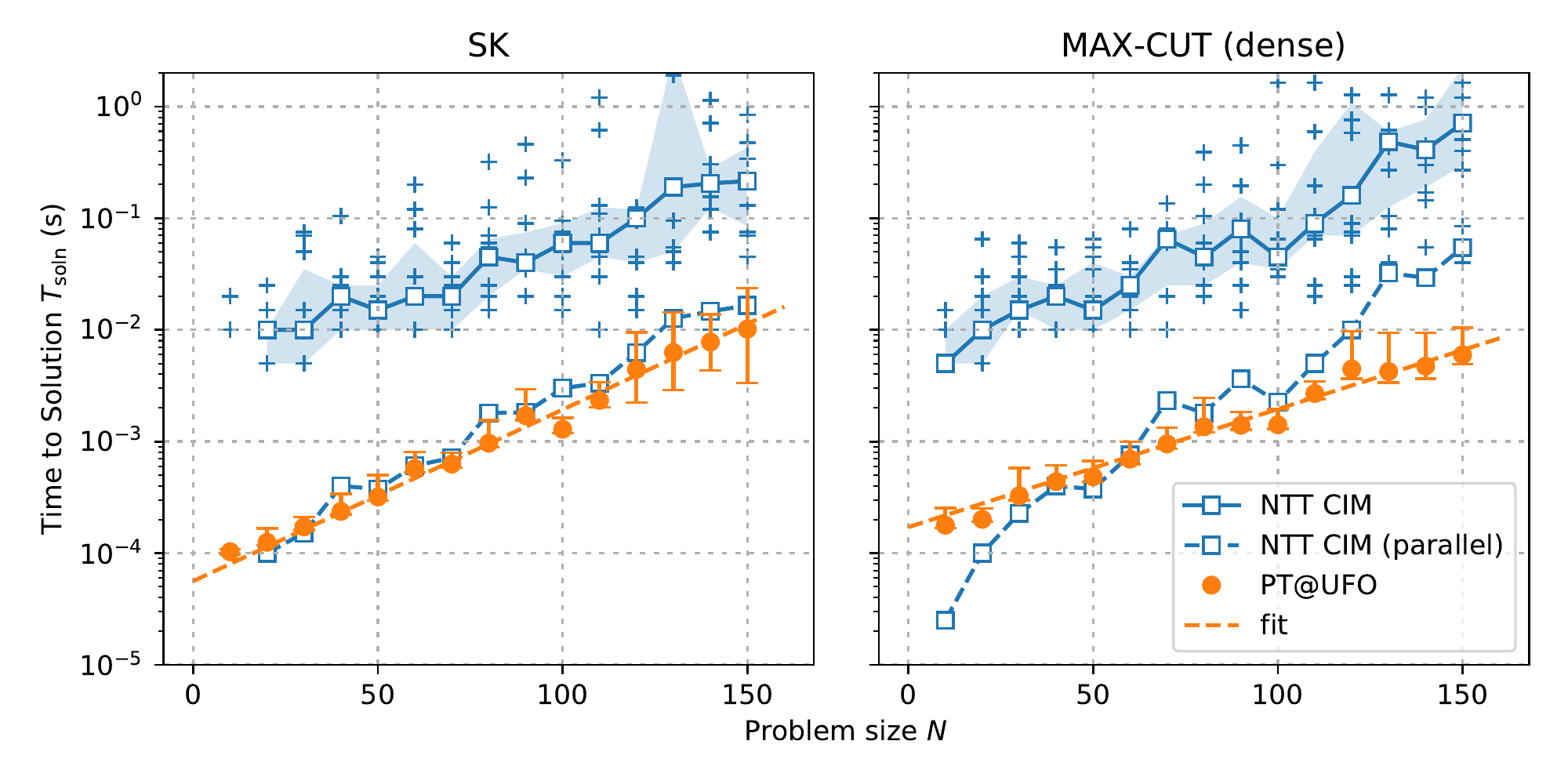}
\caption{CIM time to solution compared against the parallel tempering algorithm implemented in the Unified Framework for Optimization (UFO). The error bars for PT@UFO corresponds to the minimum and maximum value of time to solution for that specific size.  All UFO runs were performed on Intel Xeon CPU E5-1650 v2 (3.50GHz).}
\label{fig:fs12}
\end{center}
\end{figure}

\bibliographystyle{naturemag}
\bibliography{PaperRefs}

\end{document}